\documentclass[aps,12pt,draft]{revtex4}
\usepackage{latexsym}

\begin{document}

\title{Extremal Optimization for Graph Partitioning}

\author{Stefan Boettcher}
\email{sboettc@emory.edu}
\affiliation{Physics Department, Emory University, Atlanta, Georgia 30322, USA}\author{Allon G.\ Percus}\email{percus@lanl.gov}
\affiliation{Computer \& Computational Sciences Division, Los Alamos
National Laboratory, Los Alamos, NM 87545, USA}

\date{\today}
\begin{abstract}
Extremal optimization is a new general-purpose method for approximating
solutions to hard optimization problems.  We study the method in detail
by way of the NP-hard graph partitioning problem.  We discuss the
scaling behavior of extremal optimization, focusing on the convergence
of the average run as a function of runtime and system size.  The method
has a single free parameter, which we determine numerically and justify
using a simple argument.  Our numerical results demonstrate that on
random graphs, extremal optimization maintains consistent accuracy for
increasing system sizes, with an approximation error decreasing over
runtime roughly as a power law $t^{-0.4}$.  On geometrically structured
graphs, the scaling of results from the {\em average\/} run suggests
that these are far from optimal, with large fluctuations between
individual trials.  But when only the {\em best\/} runs are considered,
results consistent with theoretical arguments are recovered.
\hfil\break PACS number(s): 02.60.Pn, 05.65.+b, 75.10.Nr, 64.60.Cn.
\end{abstract}
\maketitle
\newpage

\section{Introduction}
\label{intro}

Optimizing a system of many variables with respect to some cost function
is a task frequently encountered in physics.  The determination of
ground-state configurations in disordered
materials~\cite{MPV,Grest,D+S,Disorder} and of fast-folding protein
conformations~\cite{Proteins} are but two examples.  In cases where the
relation between individual components of the system is
frustrated~\cite{Toulouse}, the cost function often exhibits a complex
``landscape''~\cite{cnls} in configuration space, posing challenges to
neighborhood search procedures.  Indeed, for growing system size the
cost function may exhibit a rapidly increasing number of unrelated local
extrema, separated by sizable barriers that can make the search for the
exact optimal solution unreasonably costly.  It is of great importance
to develop fast and reliable approximation methods for finding optimal
or acceptable near-optimal solutions with high probability.

In recent papers we have introduced a new method, called {\em extremal
optimization\/} (EO), to tackle such hard optimization
problems~\cite{BoPe1,GECCO}.  EO is a method based on the dynamics of
non-equilibrium processes and in particular those exhibiting
self-organized criticality~\cite{BTW}, where better solutions emerge
dynamically without the need for parameter tuning.  Previously, we have
discussed the basic EO algorithm, its origin, and its performance
compared with other methods.  We have demonstrated that the algorithm
can be adapted to a wide variety of NP-hard problems~\cite{PPSN_CISE}.
We have shown that for the graph partitioning problem, a simple
implementation of EO yields state-of-the-art solutions, even for systems
of $N>10^5$ variables~\cite{BoPe1}.  For large graphs of low
connectivity, EO has been shown to be faster than genetic
algorithms~\cite{Holland} and more accurate than simulated
annealing~\cite{Science}, two other widely applied methods.  A numerical
study~\cite{EOperc} has shown that EO's performance relative to
simulated annealing is particularly strong in the neighborhood of phase
transitions, ``where the really hard problems are''~\cite{Cheeseman}.
In fact, preliminary studies of the phase transition in the 3-coloring
problem~\cite{PT3COL} as well as studies of ground state configurations
in spin glasses~\cite{EO_PRL,D+S} suggest that EO may become a useful tool
in the exploration of low-temperature properties of disordered systems.

In the present work we focus on the intrinsic features of the method, by
investigating its average performance.  For this purpose, we have
performed an extensive numerical study of EO on the graph bipartitioning
problem.  We have considered various kinds of graph ensembles, both with
geometric and with random structure, for an increasing number of
vertices $N$.  The results show that for random graphs, EO converges
towards the optimal configuration in a power-law manner, typically
requiring no more than $O(N)$ update steps.  For geometric graphs the
averaged large-$N$ results are less convincing, but if we instead focus
on the best out of several trials, near-optimal results emerge.  Our
implementation of EO has one single tunable parameter, and we find a
simple relation to estimate that parameter given the allowed runtime and
system size. Many of our numerical results here have been independently 
confirmed by J.~Dall~\cite{Dall}.

The paper is organized as follows.  In Sec.~\ref{GBP} we introduce the
graph bipartitioning problem, and in Sec.~\ref{EOalgo} we describe the
extremal optimization algorithm.  Sec.~\ref{Numerics} deals in detail
with our numerical results.  In Sec.~\ref{Conclusion} we conclude with
an outlook on future work.

\section{Graph Bipartitioning}
\label{GBP}

\subsection{Definition}
The graph bipartitioning problem (GBP) is easy to formulate.  Take $N$
vertices, where $N$ is an even number, and where certain of the vertex
pairs are connected by an edge.  Then divide the vertices into two sets
of equal measure $N/2$ such that the number of edges connecting both
sets, the ``cutsize'' $m$, is minimized.  The global constraint of an
equal division of vertices places the GBP among the hardest problems in
combinatorial optimization, since determining the {\em exact\/} solution
with certainty would in general require a computational effort growing
faster than any power of $N$~\cite{NP_Hard}.  It is thus important to
find ``heuristic'' methods that can obtain good {\em approximate\/}
solutions in polynomial time.  Typical examples of applications of graph
partitioning are the design of integrated circuits (VLSI)~\cite{A+K} and
the partitioning of sparse matrices~\cite{HL}.

The general description of a graph in the previous paragraph is usually
cast in more specific terms, defining an ensemble of graphs with certain
characteristic.  These characteristics can affect the optimization
problem drastically, and often reflect real-world desiderata such as the
geometric lay-out of circuits or the random interconnections in
matrices. Therefore, let us consider a variety of different graph
ensembles, some random and some geometric in structure.

\subsection{Classes of graphs studied}
One class of graphs that has been studied extensively is that of random
graphs without geometric structure~\cite{RandGraph}.  Here, edges
between any two vertices are taken to exist with probability $p$: on the
average, an instance has a total of $E=pN(N-1)/2$ vertices and the mean
connectivity per vertex is $\alpha=p(N-1)$.  Following standard
terminology we refer to graphs of this sort as the ensemble of {\em
random graphs\/}, even though the other classes of graphs we consider
all have stochastic properties as well.

Another often-studied class of graphs without geometric structure is
generated by fixing the number of connections $\alpha$ at each vertex,
but with random connections between these vertices~\cite{W+S,Banavar}.
In particular, we consider here the case where $\alpha=3$: the ensemble
of {\em trivalent graphs\/}, randomly connected graphs with exactly
three edges originating from each vertex.

The third class we consider is an ensemble {\em with\/} geometric
structure, where the vertices are situated on a cubic lattice.  Edges
are placed so as to connect some (but not all) nearest neighbors on the
lattice: a ratio $x$ of the total number of nearest-neighboring pairs
are occupied by an edge, and those edges are distributed at random over
the possible pairs.  For a cubic lattice, the average connectivity is
then given by $\alpha=6x$.  This class of graphs corresponds to a dilute
ferromagnet, where each lattice site holds a $\pm$-spin and some (but
not all) nearest-neighboring spins possess a coupling of unit strength.
Here, the GBP amounts to the equal partitioning of $+$ and $-$ spins
while minimizing the interface between the two types~\cite{M+P}, or
simply finding the ground state under fixed (zero) magnetization.  We
thus refer to this class as the ensemble of {\em ferromagnetic
graphs\/}.

\begin{figure}
\vskip 4.0truein   
\includegraphics{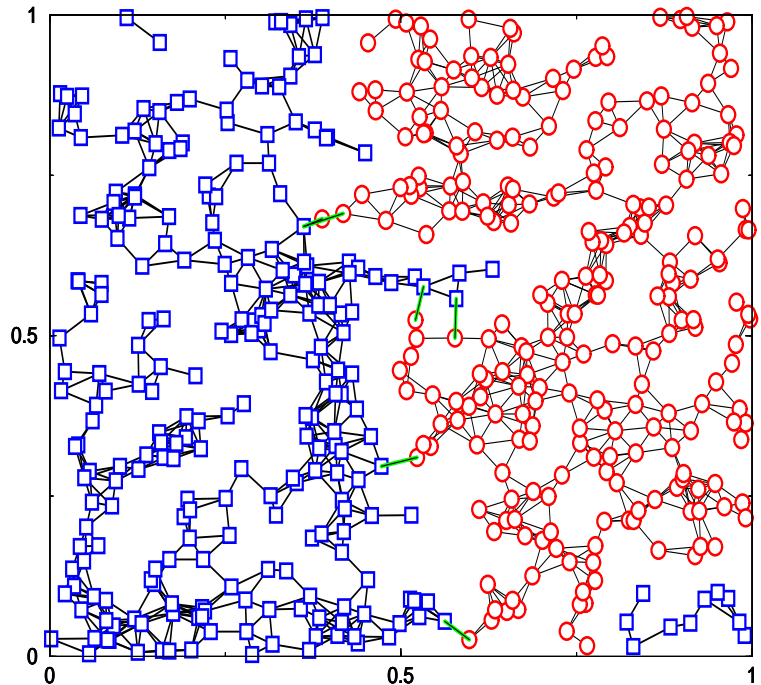} 
\caption{Plot of a geometric graph with $N=500$ vertices and average
connectivity $\alpha=6$, partitioned into 250 red and 250 blue vertices.
Starting from an initial random assignment of red and blue colors,
EO arrives at near-optimal configurations consisting of domains of
red and blue vertices, separated by an interface across which ``bad'' edges
(green lines) connect poorly-adapted vertices. }
\label{geograph}
\end{figure}

The final class we consider is that of geometric graphs specified by $N$
randomly distributed vertices in the 2-dimensional unit square, where we
place edges between all pairs whose two vertices are separated by a
distance of no more than $d$~\cite{JohnsonGBP}.  The average
connectivity is then given by $\alpha=N\pi d^2$.  The GBP on this class
of graphs has the advantage of a simple visual representation, shown in
Fig.~\ref{geograph}.  Again following standard terminology, we refer to
this class simply as the ensemble of {\em geometric graphs\/}.

It is known that graphs without geometric structure, such as those in
the first two classes, are typically easier to optimize than those with
geometric structure, such as those in the final two
classes~\cite{JohnsonGBP}.  The characteristics of the GBP for
non-geometric and geometric graphs at low connectivity appear to be very
different, due to the dominance of long loops in the former and short
loops in the latter.  The ensemble of random graphs has a structure that
is locally tree-like, allowing for a mean-field treatment that yields
some exact results~\cite{M+P}.  By contrast, the ensemble of geometric
graphs corresponds to continuum percolation of ``soft'' (overlapping)
circles, for which precise numerical results exist~\cite{Balberg}.

Each of the graph ensembles that we consider is characterized by a
control parameter, the average connectivity $\alpha$.  The difficulty of
the optimization problem for each type varies significantly with
$\alpha$.  In this study we focus on sparse graphs for which $\alpha$ is
kept constant, independent of $N$.  Sparse graphs have very different
properties from the dense graphs studied by Fu and Anderson~\cite{F+A}.
These sparse graphs are generally considered to pose the most difficult
partitioning problems, and extremal optimization (EO) is particularly
competitive in this regime~\cite{EOperc}.  In order to facilitate our
average performance study, we fix $\alpha$ to a given value on each
ensemble.  For random graphs, where the connectivity varies between
vertices according to a Poisson distribution, let $\alpha=p(N-1)=2$ (and
so $p\sim1/N$).  For trivalent graphs, by construction $\alpha=3$.  For
ferromagnetic graphs, let $\alpha=6x=2$.  For geometric graphs, let
$\alpha=N\pi d^2=6$.  In all of these cases, the connectivity is chosen
to be just above the phase transition at $\alpha_{\rm crit}$, below
which the cutsize $m$ almost always vanishes~\cite{EOperc}.  These
critical regions are especially interesting because they have been
conjectured to coincide with the hardest-to-solve instances in many
combinatorial optimization problems~\cite{Cheeseman,PhaseTrans}.

Finally, in light of the numerous comparisons in the physics literature
between the GBP and the problem of finding ground states of spin
glasses~\cite{MPV}, it is important to point out the main difference.
This is highlighted by the ensemble of ferromagnetic graphs.  Since
couplings between spins are purely ferromagnetic, all connected spins
invariably would like to be in the same state; there is no possibility
of local frustration.  Frustration in the GBP merely arises from the
{\em global\/} constraint of an equal partition, forcing spins along an
interface to attain an unfavorable state
(see Fig.~\ref{geograph}).  All other spins reside in
bulk regions where they can maintain the same state as their neighbors.
In a spin glass, on the other hand, couplings can be both ferromagnetic
and anti-ferromagnetic.  Spins everywhere have to compromise according
to conflicting conditions imposed by their neighbors; frustration is
local rather than global.

\subsection{Basic scaling arguments}
\label{scaling}
If we neglect the fact that the structure of these sparse graphs is that
of percolation clusters, we can obtain some elementary insights into the
expected scaling behavior of the optimal cutsize with increasing size
$N$, $m\sim N^{1/\nu}$.  For graphs without geometric structure (random
graph ensemble and trivalent graph ensemble), one can expect that the
cutsize should grow linearly in $N$, {\em i.e.\/}, $\nu=1$.  Indeed,
this argument can be made rigorous for arbitrary connectivity $\alpha$.
Extremal optimization performs very well on these graphs, and previous
numerical studies using EO all give $\nu\approx 1$~\cite{EOperc}.

For graphs with geometric structure (ferromagnetic graph ensemble and
geometric graph ensemble), the value of $\nu$ is less clear.  We can
approximate a graph with a $d$-dimensional geometric structure as a
hyper-cubic lattice of length $L=N^{1/d}$, where the lattice sites are
the vertices of the graph and the nearest-neighbor bonds are the edges,
of which only a finite fraction are occupied.  There are thus about
$E\sim N$ edges in the graph.  To partition the graph, we are roughly
looking for a $(d-1)$-dimensional hyper-plane cutting the graph into two
equal-sized sets of vertices.  Such an interface between the partitions
would cut $\sim L^{d-1}$ bonds, and thus $\sim N^{1-1/d}$ edges.
Following this argument, the 3-d ferromagnetic graphs
should have a cutsize scaling with $N^{2/3}$ and the 2-d geometric graphs
should have a cutsize scaling with $N^{1/2}$.

However, while this may the case for a typical partition of the graph,
it may not be the case for an {\em optimal\/} partition.  The interface
for an optimal cut of a sparse graph could well be much rougher than our
argument suggests, taking advantage of large voids between clusters of
connected vertices.  The number of cut edges would then be much below
the estimate based on assuming a flat interface, making $1/\nu < 1-1/d$.
In our previous studies using EO, we found $1/\nu\approx 0.75\pm0.05$
for ferromagnetic graphs and $1/\nu\approx 0.6\pm0.1$ for geometric
graphs~\cite{EOperc}, {\em i.e.\/}, above the upper bound, and our newer
results do not improve on these (seen later in Fig.~\ref{aver_best}).
This could
indicate that the actual values are close to the upper bound, but also
that for graphs with geometric structure EO fails to find the optima on
instances of increasing size.  

Similar behavior has been observed with other local search
methods~\cite{JohnsonGBP}, reflecting the fact that sparse geometric
graphs generally pose a much greater challenge than do sparse random
graphs. In contrast, a heuristic such as METIS~\cite{METIS}, a
hierarchical decomposition scheme for partitioning problems, works
better for geometric graphs than for random graphs~\cite{unpublished}.
METIS performs particularly well for sparse geometric graphs, and
typically produces better results than EO for $\alpha=6$.  Furthermore,
if speed is the dominant requirement, METIS is superior to any local
search method by at least a factor of $N$.  But for random graphs at
$\alpha=2$ or the trivalent graphs, METIS' results are poor compared to
EO's, and for all type of graphs METIS' performance deteriorates with
increasing connectivity.

\section{Extremal Optimization Algorithm}
\label{EOalgo}

\subsection{Motivation}
The extremal optimization method originates from insights into the
dynamics of non-equilibrium critical phenomena.  In particular, it is
modeled after the Bak-Sneppen mechanism~\cite{BS}, which was introduced
to describe the dynamics of co-evolving species.

Species in the Bak-Sneppen model are located on the sites of a lattice,
and each one has a ``fitness'' represented by a value between 0 and 1.
At each update step, the smallest value (representing the most poorly
adapted species) is discarded and replaced by a new value drawn randomly
from a flat distribution on $[0,1]$.  Without any interactions, all the
fitnesses in the system would eventually approach 1.  But obvious
interdependencies between species provide constraints for balancing the
system's overall condition with that of its members: the change in
fitness of one species impacts the fitness of an interrelated species.
Therefore, at each update step, the Bak-Sneppen model replaces the
fitness values on the sites {\em neighboring\/} the smallest value with
new random numbers as well.  No explicit definition is provided for the
mechanism by which these neighboring species are related.  Yet after a
certain number of updates, the system organizes itself into a highly
correlated state known as self-organized criticality (SOC)~\cite{BTW}.
In that state, almost all species have reached a fitness above a certain
threshold.  But these species merely possess what is called punctuated
equilibrium~\cite{G+E}: since only one's weakened neighbor can undermine
one's own fitness, long periods of ``stasis'', with a fitness above the
threshold, are inevitably punctuated by bursts of activity.  This
co-evolutionary activity cascades in a chain reaction (``avalanche'')
through the system.  These fluctuations can involve any number of
species, up to the system size, making any possible configuration
accessible.  Due to the extremal nature of the update, however, the
system as a whole will always return to states in which practically all
species are above threshold.

In the Bak-Sneppen model, the high degree of adaptation of most species
is obtained by the elimination of poorly adapted ones rather than by a
particular ``engineering'' of better ones.  While such dynamics might
not lead to as optimal a solution as could be engineered under specific
circumstances, it provides near-optimal solutions with a high degree of
latency for a rapid adaptation response to changes in the resources that
drive the system.  A similar mechanism, based on the Bak-Sneppen model,
has recently been proposed to describe adaptive learning in the
brain~\cite{C+B}.

\subsection{Algorithm description}
\label{eo_describe}
Inspired by the Bak-Sneppen mechanism, we have devised the EO algorithm
with the goal of accessing near-optimal configurations for hard
optimization problems using a minimum of external control.  Previously,
we have demonstrated that the EO algorithm is applicable to a wide
variety of problems~\cite{EO_PRL,PPSN_CISE}.  Here, we focus on its
implementation for the GBP.

In the GBP, EO~\cite{BoPe1} considers each vertex of a graph as an
individual variable with its own fitness parameter.  It assigns to each
vertex $i$ a ``fitness'' 
\begin{eqnarray}
\lambda_i=g_i/(g_i+b_i),
\label{fitnesseq}
\end{eqnarray}
where $g_i$ is the number of ``good'' edges connecting $i$ to other
vertices within its same set, and $b_i$ is the number of and ``bad''
edges connecting $i$ to vertices across the partition.  (For unconnected
vertices we fix $\lambda_i=1$.)  Note that vertex $i$ has an individual
connectivity of $\alpha_i=g_i+b_i$, while the overall mean connectivity
of a graph is given by $\alpha=\sum_i\alpha_i/N$ and the cutsize of a
configuration is given by $m=\sum_ib_i/2$.

At all times an ordered list is maintained, in the form of a permutation
$\Pi$ of the vertex labels $i$ such that
\begin{eqnarray}
\lambda_{\Pi(1)}\leq\lambda_{\Pi(2)}\leq\ldots\leq\lambda_{\Pi(N)},
\label{rankeq}
\end{eqnarray}
and $i=\Pi(k)$ is the label of the $k$th ranked vertex in the list.

Feasible configurations have $N/2$ vertices in one set and $N/2$ in the
other.  To define a local search of the configuration space, we must
define a ``neighborhood'' for each configuration within this
space~\cite{Reeves}.  The simplest such neighborhood for the GBP is
given by an exchange of (any) two spins between the sets.  With this
exchange at each update we can {\em search\/} the configuration space by
moving from the current configuration to a neighboring one.  In close
analogy with the Bak-Sneppen mechanism, our original parameter-free
implementation of EO merely swapped the vertex with the worst fitness,
$\Pi(1)$, with a random vertex from the opposite set~\cite{BoPe1}.  Over
the course of a run with $t_{\rm max}$ update steps, the cutsize of the
configurations explored varies widely, since each update can result in
better or worse fitnesses.  Proceeding as with the gap-equation in the
Bak-Sneppen model~\cite{PMB}, we can define a function $m(t)$ to be to
be the cutsize of the {\em best\/} configuration seen during this run up
to time $t$.  By construction $m(t)$ is monotonically decreasing, and
$m(t_{\rm max})$ is the output of a single run of the EO-algorithm.

We find that somewhat improved results are obtained with the following
one-parameter implementation of EO.  Draw two integers, $1\leq
k_1,k_2\leq N$, from a probability distribution
\begin{eqnarray}
P(k)\propto k^{-\tau},\quad(1\leq k\leq N),
\label{pdfeq}
\end{eqnarray}
on each update, for some $\tau$.  Then pick the vertices $i_1=\Pi(k_1)$
and $i_2=\Pi(k_2)$ from the rank-ordered list of fitnesses in
Eq.~(\ref{rankeq}). (We repeatedly draw $k_2$ until we obtain a vertex
in the opposite set from $k_1$.)  Let vertices $i_1$ and $i_2$ exchange
sets {\em no matter what\/} the resulting new cutsize may be.  Then,
reevaluate the fitnesses $\lambda$ for $i_1$, $i_2$, and all vertices
they are connected to ($2\alpha$ on average).  Finally, reorder the
ranked list of $\lambda$'s according to Eq.~(\ref{rankeq}), and start the
process over again.  Repeat this procedure for a number of update steps
per run that is linear in system size, $t_{\rm max}=AN$, and store the
best result generated along the way.  Note that no scales to limit
fluctuations are introduced into the process, since the selection
follows the scale-free power-law distribution $P(k)$ in Eq.~(\ref{pdfeq})
and since --- unlike in heat bath methods --- all moves are accepted.
Instead of a global cost function, the rank-ordered list of fitnesses
provides the information about optimal configurations.  This information
emerges in a self-organized manner merely by selecting with a bias {\em
against\/} badly adapted vertices, rather than ever ``breeding'' better
ones.

\subsection{Discussion}
\subsubsection{Definition of fitness}
We now discuss some of the finer details of the algorithm.  First of
all, we stress that 
we use the term ``fitness'' in the sense of the
Bak-Sneppen model, in marked contrast to its meaning in
genetic algorithms.  Genetic algorithms consider a population of
configurations and assign a fitness value to an entire configuration.
EO works with only a single configuration and makes local updates to
individual variables within that configuration.  Thus, it is important
to reiterate that EO assigns a fitness $\lambda_i$ to each of the
system's $N$ variables, rather than to the system as a whole.

While the definition of fitness in Eq.~(\ref{fitnesseq}) for the graph
partitioning problem seems natural, it is by no means unique.  In fact,
in general the sum of the fitnesses does not even represent the cost
function we set out to optimize, because each fitness is locally
normalized by the total number of edges touching that vertex.  It may
seem more appropriate to define fitness instead as $\lambda_i=g_i$, the
number of ``good'' connections at a vertex, or else as $\lambda_i=-b_i$,
which amounts to penalizing a vertex for its number of ``bad''
connections.  In both cases, the sum of all the fitnesses is indeed
linearly related to the actual cost function.  The first of these
choices leads to terrible results, since almost all vertices in
near-optimal configurations have only good edges, and so in most cases
$g_i$ is simply equal to the connectivity of the vertex.  The second
choice does lead to a viable optimization procedure and one that is
easily generalizable to other problems, as we have shown
elsewhere~\cite{PPSN_CISE}.  But for the GBP, we find that the results
from that fitness definition are of poorer quality than those we present
in this paper.  It appears productive to consider all vertices in the
GBP on an equal footing by normalizing their fitnesses by their
connectivity as in Eq.~(\ref{fitnesseq}), so that $\lambda_i\in[0,1]$.
The fact that each vertex's pursuit towards a better fitness simultaneously
minimizes its own contribution to the total cost function ensures that EO
always returns sufficiently close to actual minima of the cost function.

Note that ambiguities similar to that of the fitness
definition also occur for other optimization methods.  In general,
there is usually a large variety of different neighborhoods to
choose from.  Furthermore, to facilitate a local neighborhood search,
cost functions often have to be amended to contain penalty terms.  It
has long been known~\cite{JohnsonGBP} that simulated annealing
for the GBP only becomes effective when one allows the balanced partition
constraint to be violated, using an extra term in the cost function to
represent unbalanced partitions in the cost function.  Controlling this
penalty term requires an additional parameter and additional tuning,
which EO avoids.

\subsubsection{The parameter $\tau$}
\label{estimate}
Indeed, there is only one parameter, the exponent $\tau$ in the
probability distribution in Eq.~(\ref{pdfeq}), governing the update
process and consequently the performance of EO.  It is intuitive that a
value of $\tau$ should exist that optimizes EO's average-case
performance.  If $\tau$ is too small, vertices would be picked purely at
random with no gradient towards any good partitions.  If $\tau$ is too
large, only a small number of vertices with particularly bad fitness
would be chosen over and over again, confining the system to a poor
local optimum.  Fortunately, we can derive an asymptotic relation that
estimates a suitable value for $\tau$ as a function of the allowed
runtime and the system size $N$.  The argument is actually {\em
independent\/} of the optimization problem under consideration, and is
based merely on the probability distribution in Eq.~(\ref{pdfeq}) and
the ergodicity properties that arise from it.

We have observed numerically that the choice of an optimal $\tau$
coincides with the transition of the EO algorithm from ergodic to
non-ergodic behavior.  But what do we mean by ``ergodic'' behavior, when
we are not in an equilibrium situation?  Consider the rank-ordered list
of fitnesses in Eq.~(\ref{rankeq}), from which we choose the individual
variables to be updated.  If $\tau=0$, we choose variables at random and
can reach every possible configuration of the system with equal
probability.  EO's behavior is then perfectly ergodic.  Conversely, if
$\tau$ is very large, there will be at least a few variables that may
never be touched in a finite runtime $t$, because they are already
sufficiently fit and high in rank $k$.  Hence, if there are
configurations of the system that can only be reached by altering these
variables first, EO will never explore them and accordingly is
non-ergodic.  Of course, for any finite $\tau$, different configurations
will be explored by EO with different probabilities.  But we argue that
phenomenologically, we may describe EO's behavior as ergodic provided
every variable, and hence every rank on the list, gets selected at least
once during a single run.  There will be a value of $\tau$ at which
certain ranks, and therefore certain variables, will no longer be
selected with finite probability during a given runtime.  We find that
this value of $\tau$ at the transition to non-ergodic behavior
corresponds roughly with the value at which EO displays its best
performance.  Clearly, this makes the choice of $\tau$ dependent on the
runtime and on the system size $N$.

Assuming that this coincidence indicates a causal relation between the
ergodic transition and optimal performance, we can estimate the optimal
$\tau$ in terms of runtime $t$ and size $N$.  EO uses a runtime $t_{\rm
max}=AN$, where $A$ is a constant, probably much larger than 1 but much
smaller than $N$.  (For a justification of the runtime scaling linearly
in $N$, see Sec.~\ref{randscal}.) We argue that we are at the
``edge of ergodicity'' when during $AN$ update steps we have a chance of
selecting even the highest rank in EO's fitness list, $k=N$, about once:
\begin{eqnarray}
P(k=N)AN\sim1.
\end{eqnarray}
With the choice of the power-law distribution for $P(k)$ in
Eq.~(\ref{pdfeq}), we obtain
\begin{eqnarray}
(\tau-1)N^{-\tau}AN\sim1,\quad(N\to\infty),
\end{eqnarray}
where the factor of $\tau-1$ arises from the norm of $P(k)$. Asymptotically,
we find
\begin{eqnarray}
\tau\sim1+{\log\left(A/\log N\right)\over\log N}\quad
(N\to\infty,1\ll A\ll N).
\label{tauscaleq}
\end{eqnarray}
Of course, large higher-order corrections may exist, and there may well
be deviations in the optimal $\tau$ among different classes of graphs
since this argument does not take into account graph structure
or even the problem at hand.
Nevertheless, Eq.~(\ref{tauscaleq}) gives a qualitative understanding of
how to choose $\tau$, indicating for instance that it varies very slowly
with $N$ and $A$ but will most likely be significantly larger than its
asymptotic value of unity.  Not surprisingly, with the numerical values
$A\approx10^2$ and $N\approx10^4$ used in previous studies, we
typically have observed optimal performance for $\tau\approx1.3-1.6$ 
(see also Ref.~\cite{Dall}). Our
numerical study of $\tau$ is discussed in Sec.~\ref{ergo} below.
(We note that in this study we often use runtimes with $A>N$ to probe the
extreme long-time convergence behavior of EO. In that case,
Eq.~(\ref{tauscaleq}) can not be expected to apply. Yet, the optimal 
value of $\tau$ still increases with $A$, as will be seen in the numerical
results in Sec.~\ref{ergo} and Fig.~\ref{logscal_glass}a.)

\subsubsection{Efficient ranking of the fitness values}  
Strictly speaking, the EO-algorithm as we have described it has a cost
proportional to $2\alpha N^2\log N$ per run.  One factor of $N$ arises
simply from the fact that the runtime, {\em i.e.\/}, the number of
update steps per run, is taken to scale linearly with the system size.
The remaining factor of $2\alpha N\log N$ arises from the necessity to
maintain the ordered list of fitnesses in Eq.~(\ref{rankeq}): during
each update, on average $2\alpha$ vertices change their fitnesses and
need to be reordered, since the two vertices chosen to swap partitions
are each connected on average to $\alpha$ other vertices.  The cost of
sequentially ordering fitness values is in principle $N\log N$.
However, to save a factor of $N$, we have instead resorted to an
imperfect heap ordering of the fitness values, as already described in
Ref.~\cite{BoPe1}.  Ordering a list of $N$ numbers in a binary tree or
``heap'' ensures that the smallest fitness will be at the root of the
tree, but does not provide a perfect ordering between all of the other
members of the list as a sequential ordering would.  Yet, with high
probability smaller fitnesses still reside in levels of the tree closer
to the root, while larger fitnesses reside closer to the end-nodes of
the tree.  To maintain such a tree only requires the $O(\log N)$ moves
needed to replace changing fitness values.

Specifically, consider a list of $N$ fitness values. This list will fill
a binary tree with at most $l_{\rm max}+1$ levels, where $l_{\rm
max}=[\log_2(N)]$ ($[x]$ denotes the integer part of $x$) and
$l=0,1,\ldots,l_{\rm max}$, where $l=0$ is the level consisting solely
of the root, $l=1$ is the level consisting of the two elements extending
from the root, etc.  In general, the $l$th level contains up to $2^l$
elements, and all level are completely filled except for the
end-node-level $l_{\rm max}$, which will only be partially filled in
case that $N<2^{l_{\rm max}+1}-1$.  Clearly, by definition, every
fitness on the $l$th level is worse than its two descendents on the
$(l+1)$th level, but there could nevertheless be fitnesses on the $l$th
level that are better than some of the other fitnesses on the $(l+1)$th
(or even greater) level.  On {\em average\/}, though, the fitnesses on
the $l$th level are always worse than those on the $(l+1)$th level, and
better than those on the $(l-1)$th level.

Thus, instead of applying the probability distribution in
Eq.~(\ref{pdfeq}) directly to a sequentially ordered list, we save an
entire factor of $N$ in the computational cost by using an analogous
(exponential) probability distribution on the (logarithmic) number of
levels $l$ in our binary tree,
\begin{eqnarray}
Q(l)\propto 2^{-(\tau-1)l},\quad(0\leq l\leq [\log_2(N)]+1).
\label{pdf2eq}
\end{eqnarray}
{}From that level $l$ we then choose one of the $2^l$ fitnesses at
random and update the vertex associated with that fitness value.
Despite the differences in the implementations, our studies show that
heap ordering and sequential ordering produce quite similar results.
In particular, the optimal value of $\tau$ found for both methods is
virtually indistinguishable.  But with the update procedure using the
heap, EO algorithm runs at a cost of merely $O(2\alpha N\log N)$.

Under some circumstances, it may be possible to maintain a partially or
even perfectly ordered list at constant cost, by using a hash table.
For instance, for trivalent graphs or more generally for $\alpha$-valent
graphs, each vertex $i$ can be in only one of $\alpha+1$ attainable
states, $b_i=0,1,\ldots,\alpha$ and $\lambda_i=b_i/\alpha$. Thus,
instead of time consuming comparisons between $\lambda$'s, fitness
values can be hashed into and retrieved from ``buckets'', each
containing all vertices with a given $b_i$.  For an update, we then
obtain ranks according to Eq.~(\ref{pdfeq}), determine which bucket that
rank points to, and retrieve one vertex at random from that bucket. 

Even in cases where the fitness values do not fall neatly into a
discrete set of states, such a hash table may be an effective
approximation.  But great care must be taken with respect to the
distribution of the $\lambda$'s.  This distribution could look
dramatically different in an average configuration and in a near-optimal
configuration, because in the latter case fitness values may be densely
clustered about 1.

\subsubsection{Startup routines}
\label{startup}
The results obtained during a run of a local search algorithm can often
be refined using an appropriate startup routine.  This is an issue of
practical importance for any optimization method~\cite{Reeves}.  For
instance, Ref.~\cite{JohnsonGBP} has explored improvements for the
partitioning of geometric graphs by initially dividing the vertices of
the graph into two geometrically defined regions (for instance, drawing
a line through the unit square).  This significantly boosted the
performance of the Kernighan-Lin algorithm on the graphs.  Such methods
are not guaranteed to help, however: simulated annealing shows little
improvement using a startup~\cite{JohnsonGBP}.  Happily, the performance
of the EO algorithm typically {\em is\/} improved considerably with a clever
startup routine.

Previously, we have explored a startup routine using a simple clustering
algorithm~\cite{BoPe1}, which accomplishes the separation of the graph
into domains not only for geometric but also for random graphs.  The
routine picks a random vertex on the graph as a seed for a growing
cluster, and recursively incorporates the boundary vertices of the
cluster until $N/2$ vertices are covered, or until no boundary vertices
exist anymore (signaling a disconnected cluster in the graph).  The
procedure then continues with a new seed among the remaining vertices.
Such a routine can substantially enhance EO's convergence especially for
geometrically defined graphs (see Sec.~\ref{geomscal}).  In this paper,
though, we did {\em not\/} use any startup procedure because we prefer
to focus on the intrinsic features of the EO algorithm itself.  Instead,
all the results presented here refer to runs starting from random
initial partitions, except for a small comparison in
Sec.~\ref{geomscal}.

\section{Numerical Results}
\label{Numerics}

\subsection{Description of EO runs}
In our numerical simulations, we considered the four classes of graphs
introduced in Sec.~\ref{GBP}.  For each class, we have studied the
performance of EO as a function of the size of the graph $N$, the
runtime $t$, and the parameter $\tau$.  To obtain a statistically
meaningful test of the performance of EO, large values of $N$ were
chosen; EO performed too well on smaller graphs.  The maximum value of
$N$ varied with each kind of graph, mostly due to the fact that some
types required averaging over many more instances, thus limiting the
attainable sizes.

The precise instance sizes were: $N=1022$, 2046, 4094 and 8190 for both
random and trivalent graphs; $N=13^3(=2197)$, $16^3(=4096)$ and
$20^3(=8000)$ for ferromagnetic graphs; and $N=510$, 1022, and 2046 for
geometric graphs.  For each class of graphs, we generated a sizable
number of instances at each value of $N$: 32 for random and trivalent
graphs, 16 for ferromagnetic graphs, and 64 for geometric graphs.  For
each instance, we conducted a certain number of EO runs: 8 for random
and trivalent graphs, 16 for ferromagnetic graphs, and 64 for geometric
graphs.  Finally, the runtime (number of update steps) used for each run was
$t_{\rm max}=AN$, with $A=512$ for random graphs, $A=4096$ for trivalent
graphs, $A=1000$ for ferromagnetic graphs and $A=2048$ for geometric
graphs.

\subsection{Choosing $\tau$}
\label{ergo}
\begin{figure}
\vskip 4.6truein   \includegraphics{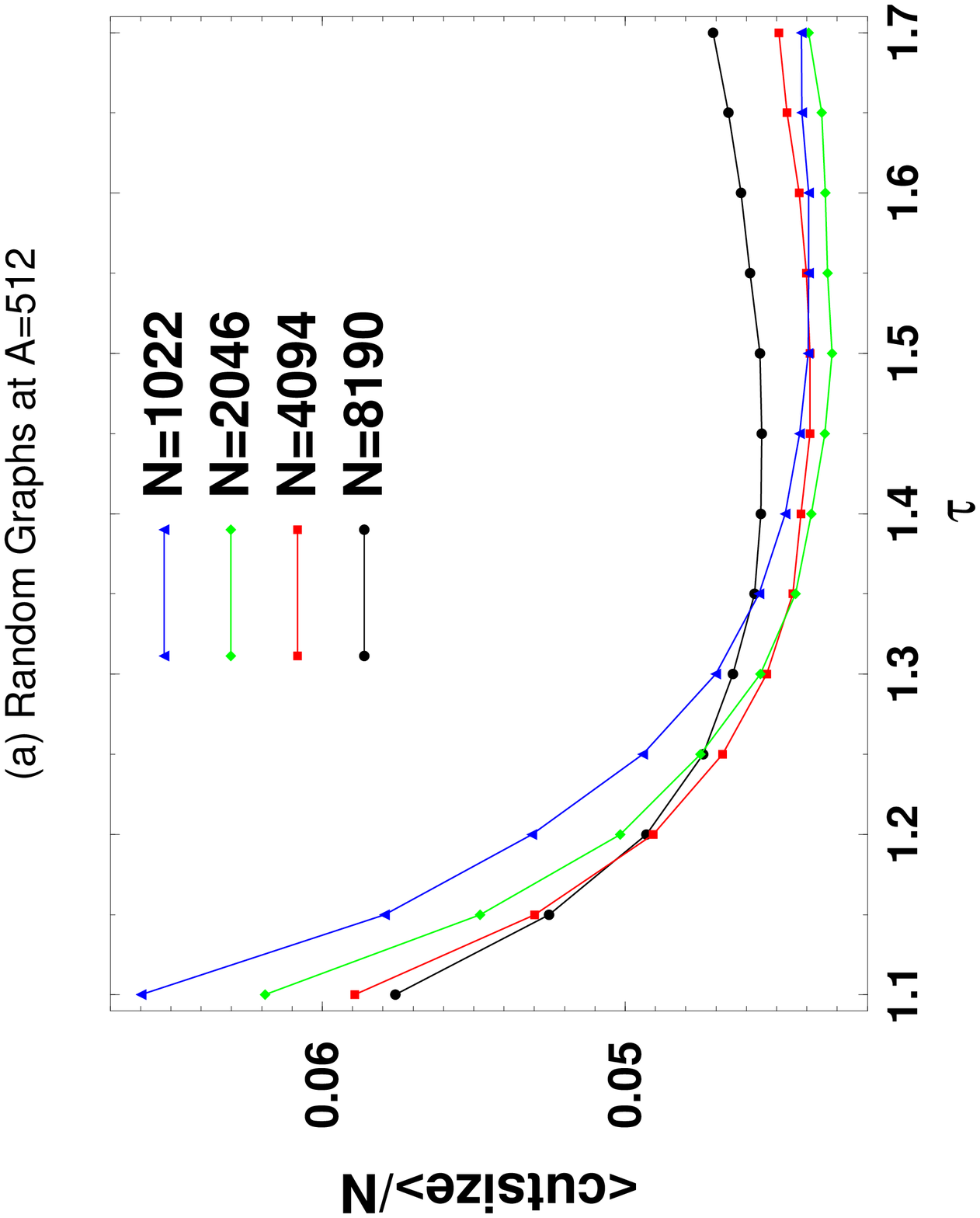}   \includegraphics{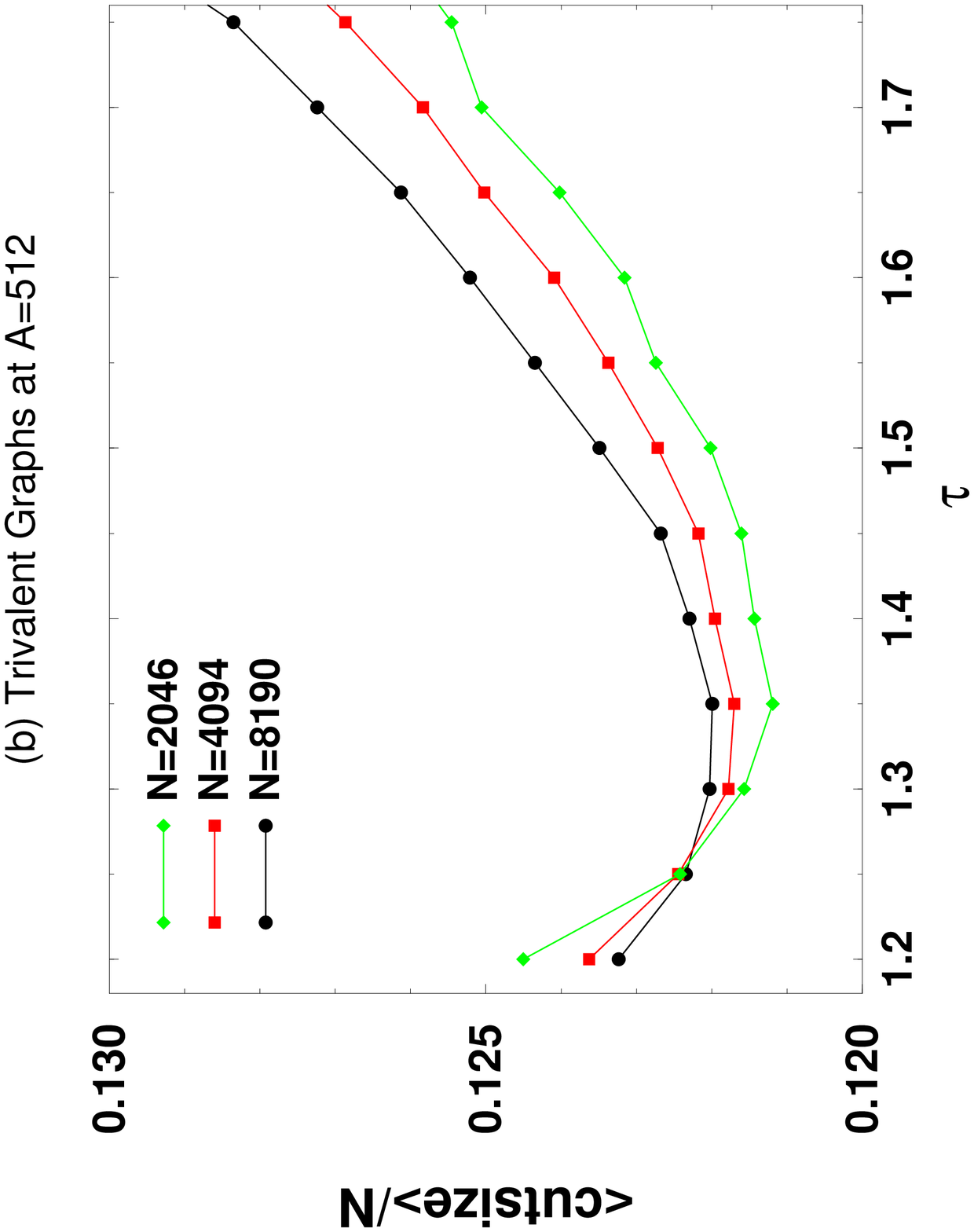}
\includegraphics{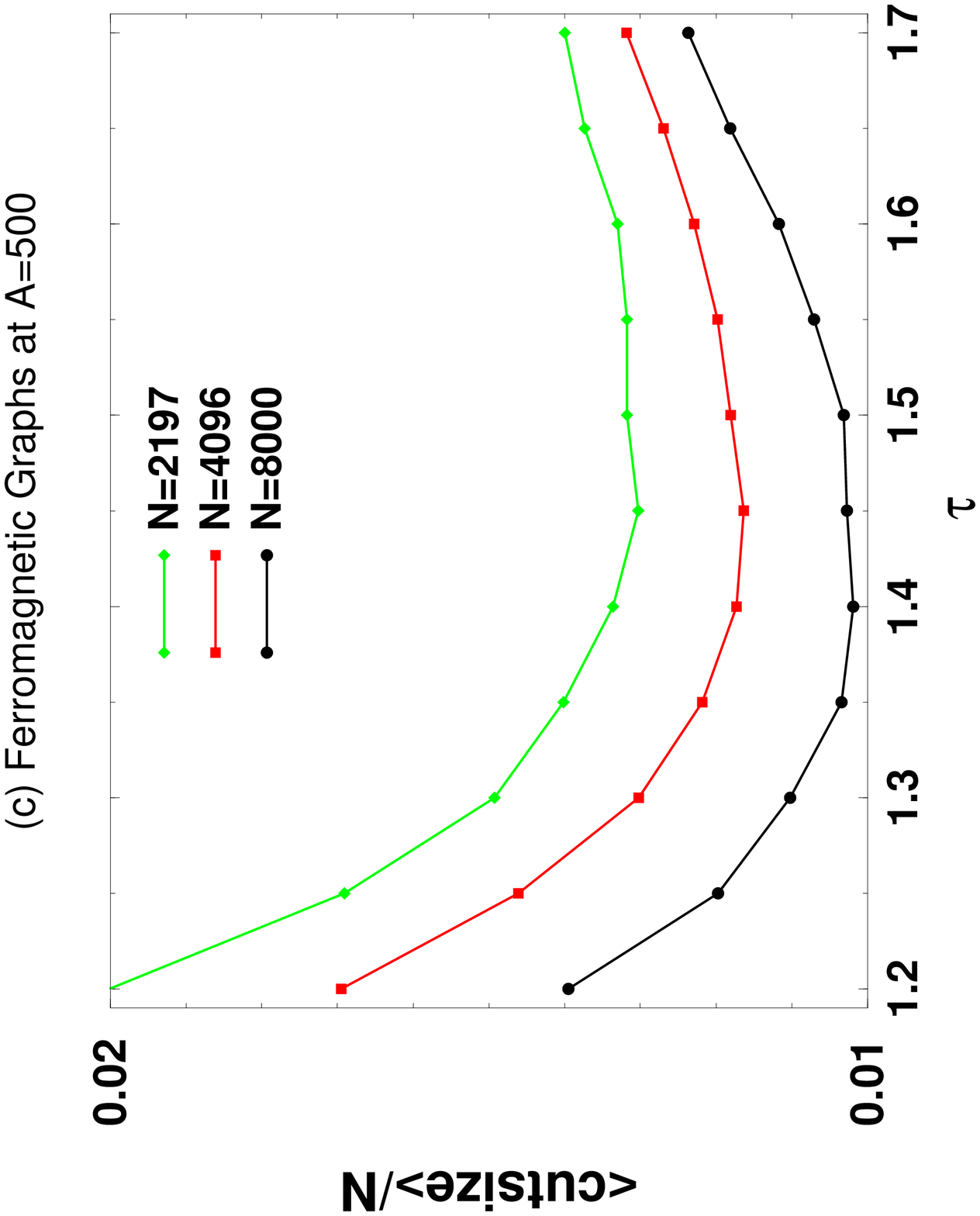}  \includegraphics{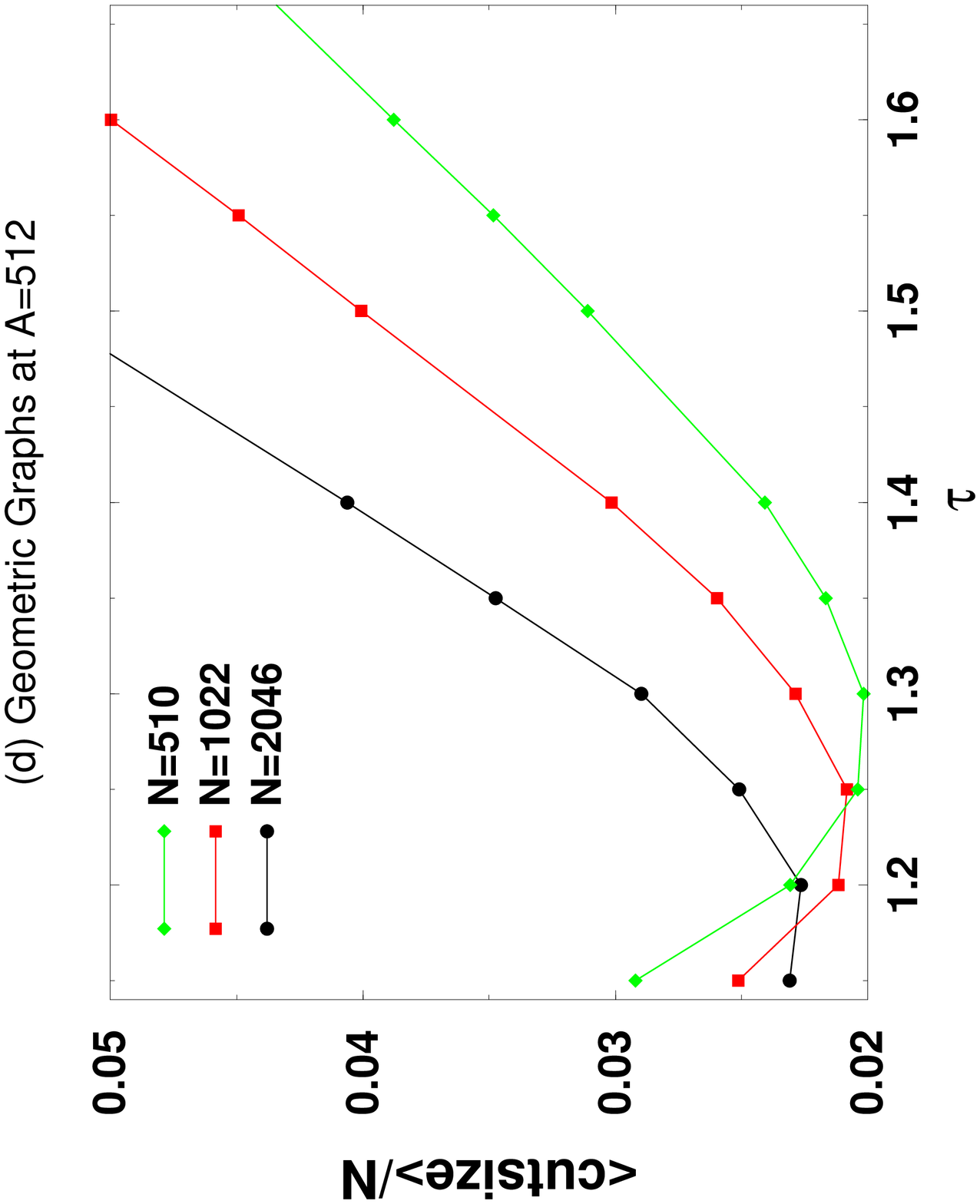}
\caption{Cutsize found by EO (in units of $N$) as a function of $\tau$
at a fixed value of $A=t/N\approx 500$, averaged over all runs on all
graphs of a given size $N$.  Results are shown for (a) random graphs,
(b) trivalent graphs, (c) ferromagnetic graphs, and (d) geometric
graphs.  For each type of graph the minimum shifts very slowly to
smaller values of $\tau$ for increasing $N$.}
\label{taufig}
\end{figure}

In previous studies, we had chosen $\tau=1.4$ as the optimal value for
all graphs.  In light of our discussion in Sec.~\ref{estimate} on how to
estimate $\tau$, here we have performed repeated runs over a range of
values of $\tau$ on the instances above, using identical initial
conditions.  The goal is to investigate numerically the optimal value
for $\tau$, and the dependence of EO's results on the value chosen.  In
Figs.~\ref{taufig}a-d we show how the average cutsize depends on $\tau$,
given fixed runtime.  There is a remarkable similarity in the
qualitative performance of EO as a function of $\tau$ for all types and
sizes of graphs.  Despite statistical fluctuations, it is clear that
there is a distinct minimum for each class, and that as expected, the
results get increasingly worse in all cases for smaller as well as
larger values of $\tau$.

\begin{figure}
\vskip 2.3truein   \includegraphics{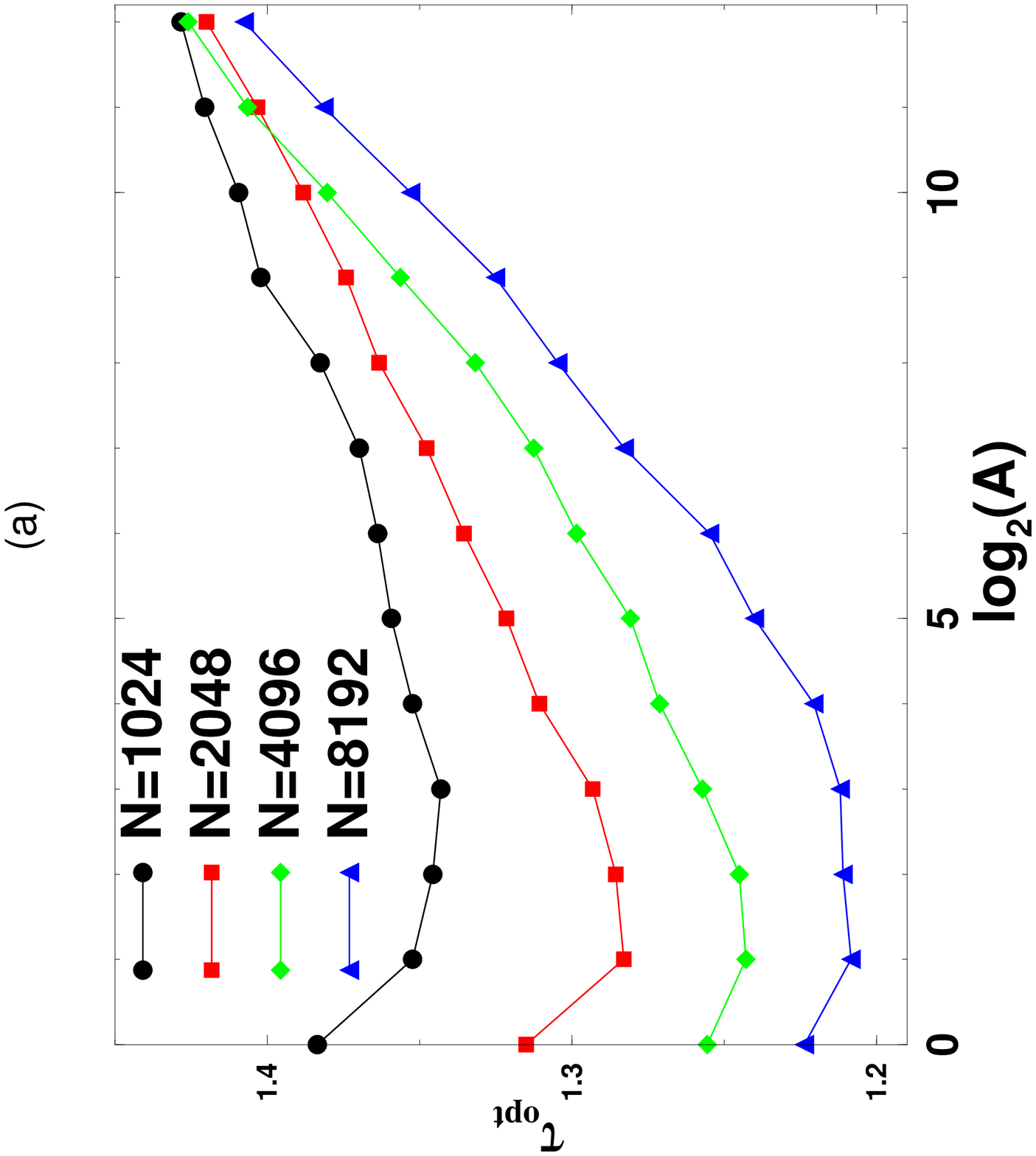}
\includegraphics{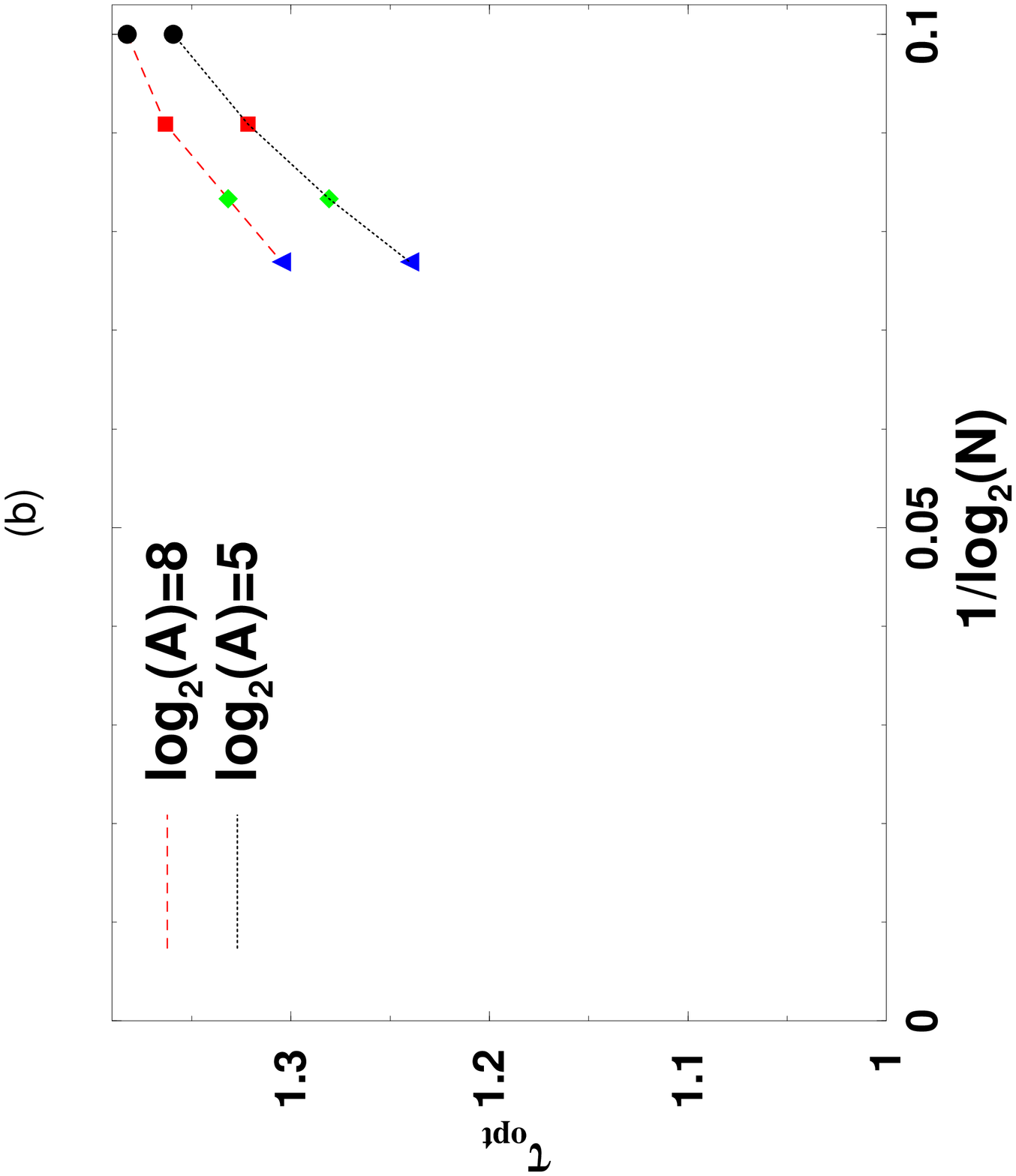}
\caption{Plot of $\tau_{\rm opt}$ for trivalent graphs as a function of
(a) $\log_2(A)=\log_2(t/N)$ for various fixed values of $N$ and (b)
$1/\log_2(N)$ for various fixed values of $A=t/N$.  These data points
were determined by performing runs as in Fig.~\protect\ref{taufig}b, and
finding the minimum of a quartic fit to data obtained at 
$\tau=1.2,~1.25,~1.3,\ldots,~1.95$.  In (a) the data increase roughly
linearly with $\log_2(A)$ in the regime $1\ll A\ll N$ at fixed $N$,
while in (b) the data extrapolate roughly toward $\tau=1$ for
$N\to\infty$ at fixed $A$, both in accordance with
Eq.~\protect\ref{tauscaleq}.  For $A\lesssim 1$ 
this scaling appears to break down, while for $A\gtrsim N$ the linear scaling
happens to remain valid.  Note that the values of $\tau_{\rm opt}$
in (b) correspond to the data points in (a) for $\log_2(A)=5$ and 8.}
\label{logscal_glass}
\end{figure}

While the optimal values for $\tau$ are similar for all types of graphs,
there is a definite drift towards smaller values of $\tau$ for
increasing $N$.  Studies on spin glasses~\cite{EO_PRL} 
have led to similar observations, supporting the argument for the
scaling of $\tau$ that we have given in Sec.~\ref{estimate}.  Our data
here do not cover a large enough range in $N$ to analyze in detail the
dependence of $\tau$ on $\log N$.  However, we can at least demonstrate
that the results for trivalent graphs, where statistical errors are
relatively small, are consistent with Eq.~(\ref{tauscaleq}) above.  For
fixed $N$ but increasing values of $A=t/N$, we see in
Fig.~\ref{logscal_glass}a that the optimal value of $\tau$ appears to
increase linearly as a function of $\log A$ in the regime $1\ll A\ll N$.
At the same time, fixing $A$ and increasing $N$, we see in
Fig.~\ref{logscal_glass}b that the optimal value of $\tau$ appears to
decrease linearly as a function of $1/\log N$, towards a value near
$\tau=1$.  Thus, for $1\ll A\ll N$, $\tau$ seems to be converging very
slowly towards 1, with a finite $N$ correction of $\sim\log A/\log N$.
This is in accordance with our estimate, Eq.~(\ref{tauscaleq}), discussed
earlier in
Sec.~\ref{estimate}. (Note that at $A\approx500$ in Figs.~\ref{taufig}a-b,
the optimal values for random and trivalent graphs are close to
$\tau\approx1.3$, consistent with Eq.~(\ref{tauscaleq}). In contrast,
in our long-time studies below, $A\gtrsim N$ and a value of $\tau=1.45$
seems preferable, consistent with Fig.~\ref{logscal_glass}a.)

\begin{figure}
\vskip 2.3truein   \includegraphics{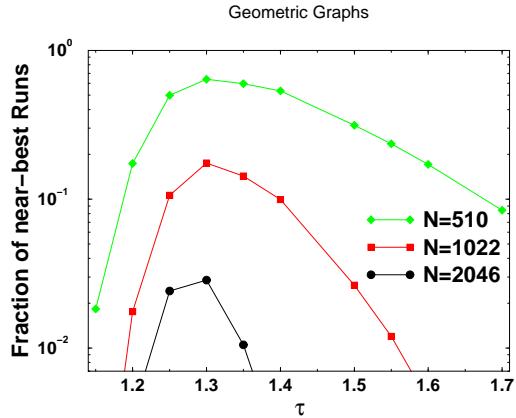} 
\caption{Fraction of EO runs on geometric graphs that have come within
20\% of the best ever found (for each instance) as a function of $\tau$,
for each value of $N$.
Maxima indicate optimal choice of $\tau$ for finding a few good results
among many runs.  These maxima occur at higher values of $\tau$ than the
minima corresponding to best average performance in
Fig.~\protect\ref{taufig}.  As seen below in
Fig.~\protect\ref{aver_best}, it becomes increasingly hard to come close
to these best values.}
\label{taufraction}
\end{figure}

The foregoing data, including the results plotted in
Figs.~\ref{taufig}a-d, arise from averages ${\overline {\langle
m(t)\rangle}}$ over all runs (denoted by $\langle\ldots\rangle$) and all
instances (denoted by an overbar).  But it is important to note that the
conclusions drawn with respect to the optimal value of $\tau$ from these
plots are valid only if there is little difference between the average
run $\langle m(t)\rangle$ and the best run $m_{\rm best}$ for a
particular instance.  While this is the case for the random and
trivalent graphs, there is a significant difference between $\langle
m(t)\rangle$ and $m_{\rm best}$ for instances of ferromagnetic and
geometric graphs, as is discussed later in Sec.~\ref{geomscal}.  In
fact, for geometric graphs (Fig.~\ref{aver_best} below), average and
best cutsizes often differ by a factor of 2 or 3!  Fig.~\ref{taufig}d
indicates that the optimal value of $\tau$ for the average performance
on geometric graphs of size $N=2046$ is below $\tau=1.2$.  If we instead
plot the fraction of runs that have come, say, within 20\% of the best
value $m_{\rm best}$ found by EO (which most likely is still not
optimal) for each instance, the optimal choice for $\tau$ shifts to
larger values, $\tau\approx1.3$ (see Fig.~\ref{taufraction}).

We may interpret this discrepancy as follows.  At lower values of
$\tau$, EO is more likely to explore the basin of many local minima but
also has a smaller chance of descending to the ``bottom'' of any basin.
At larger values of $\tau$, all but a few runs get stuck in the basin of
a poor local minimum (biasing the average) but the lucky ones have a
greater chance of finding the better states.  For geometric graphs, the
basins seem particularly hard to escape from, except at the very low
values of $\tau$ where EO is unlikely to move toward the basin's
minimum.  Thus, in such cases we find that we get better average
performance $\langle m(t)\rangle$ at a lower value of $\tau$, but the
best result $m_{\rm best}$ of multiple runs is obtained at a higher value of $\tau$.

Clearly, in order to obtain optimal performance of EO within a given
runtime and for a given class of graphs, further study of the best choice
of $\tau$ would be justified.  But for an analysis of the scaling properties
of EO with runtime and $N$, the specific details of how the optimal $\tau$
varies are not significant.  Therefore, to simply the following scaling
discussion, we will simply fix $\tau$ to a near-optimal value on each
type of graph.

\subsection{Scaling behavior}
\subsubsection{General results}
\begin{figure}[!t]
\vskip 4.6truein   \includegraphics{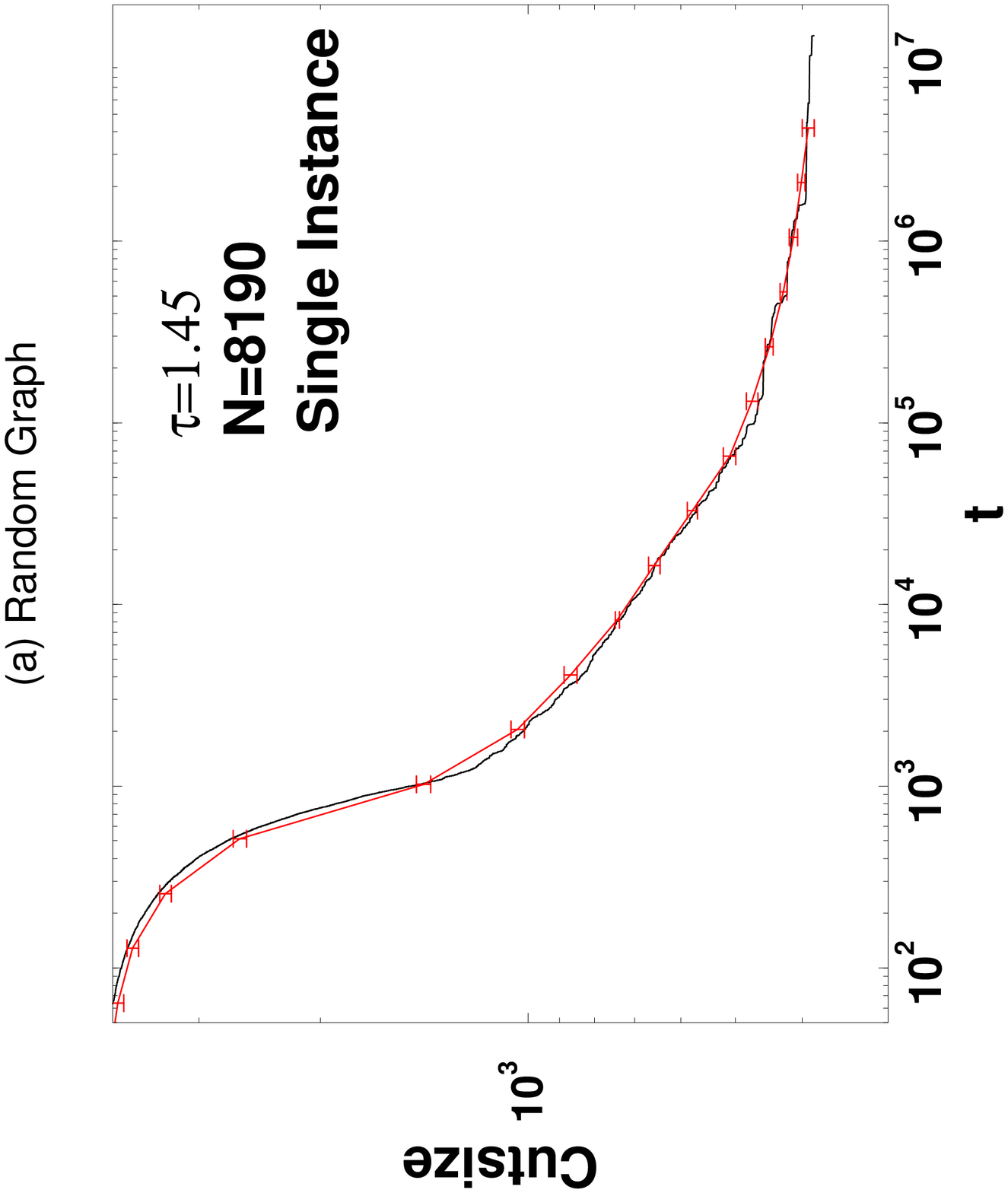} \includegraphics{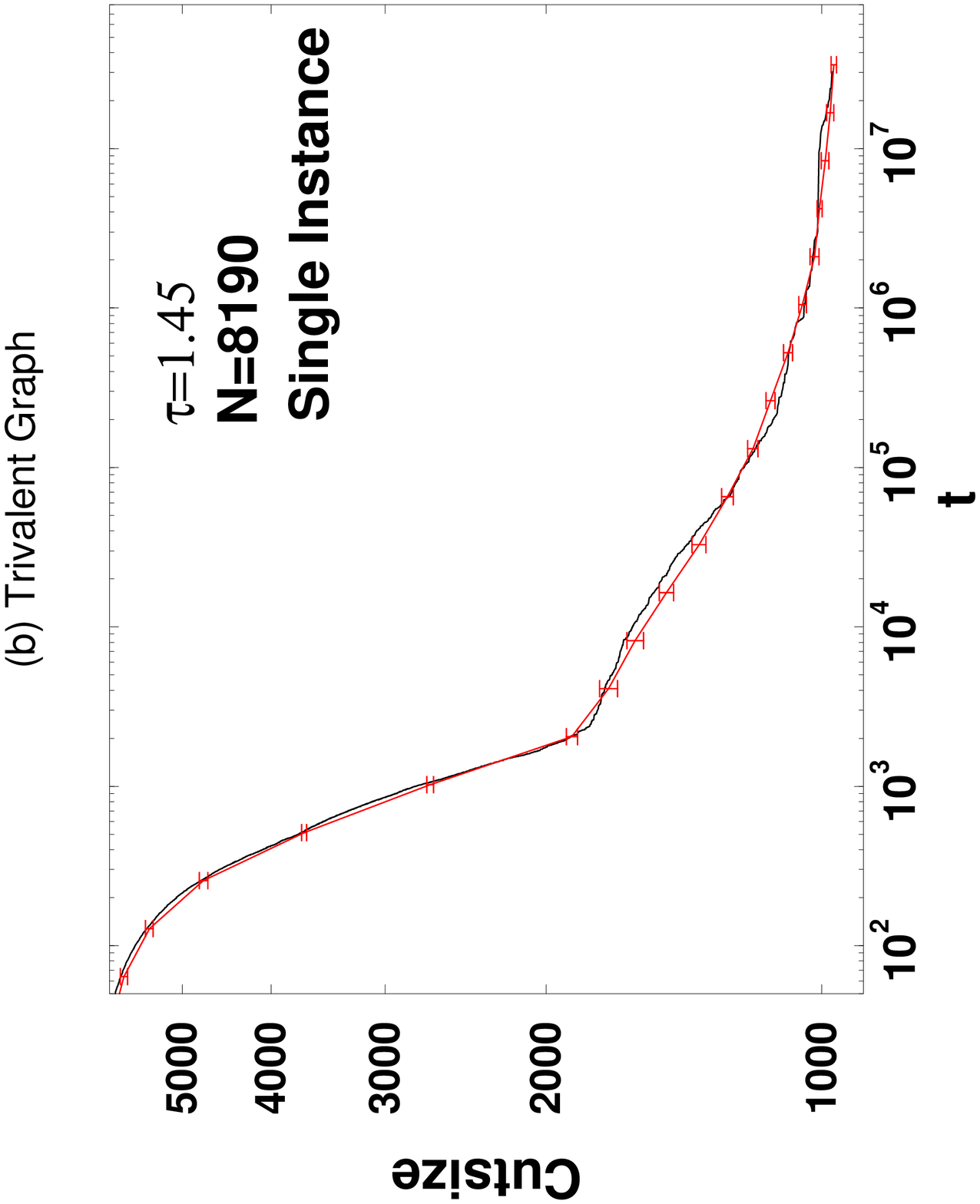}
\includegraphics{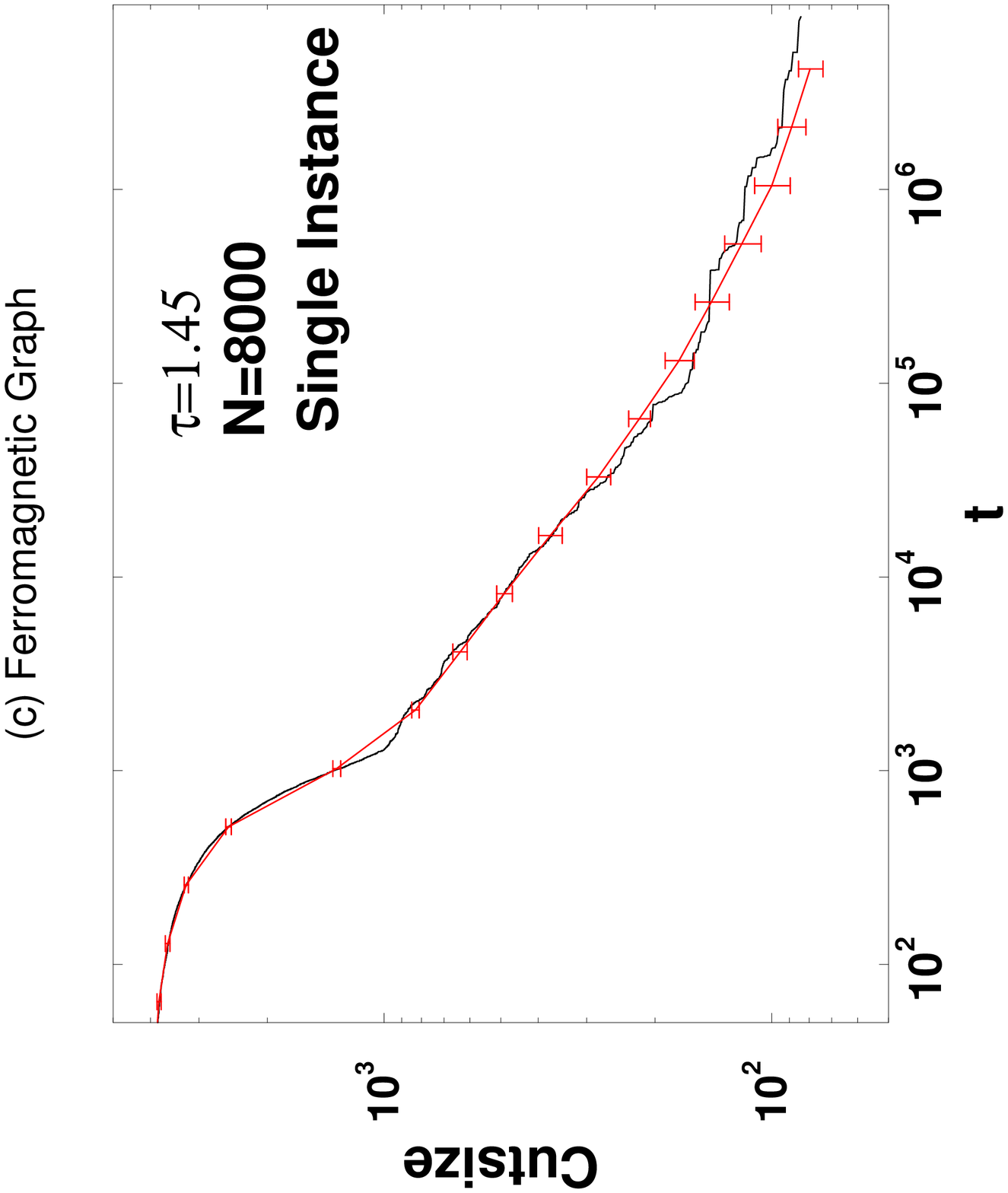} \includegraphics{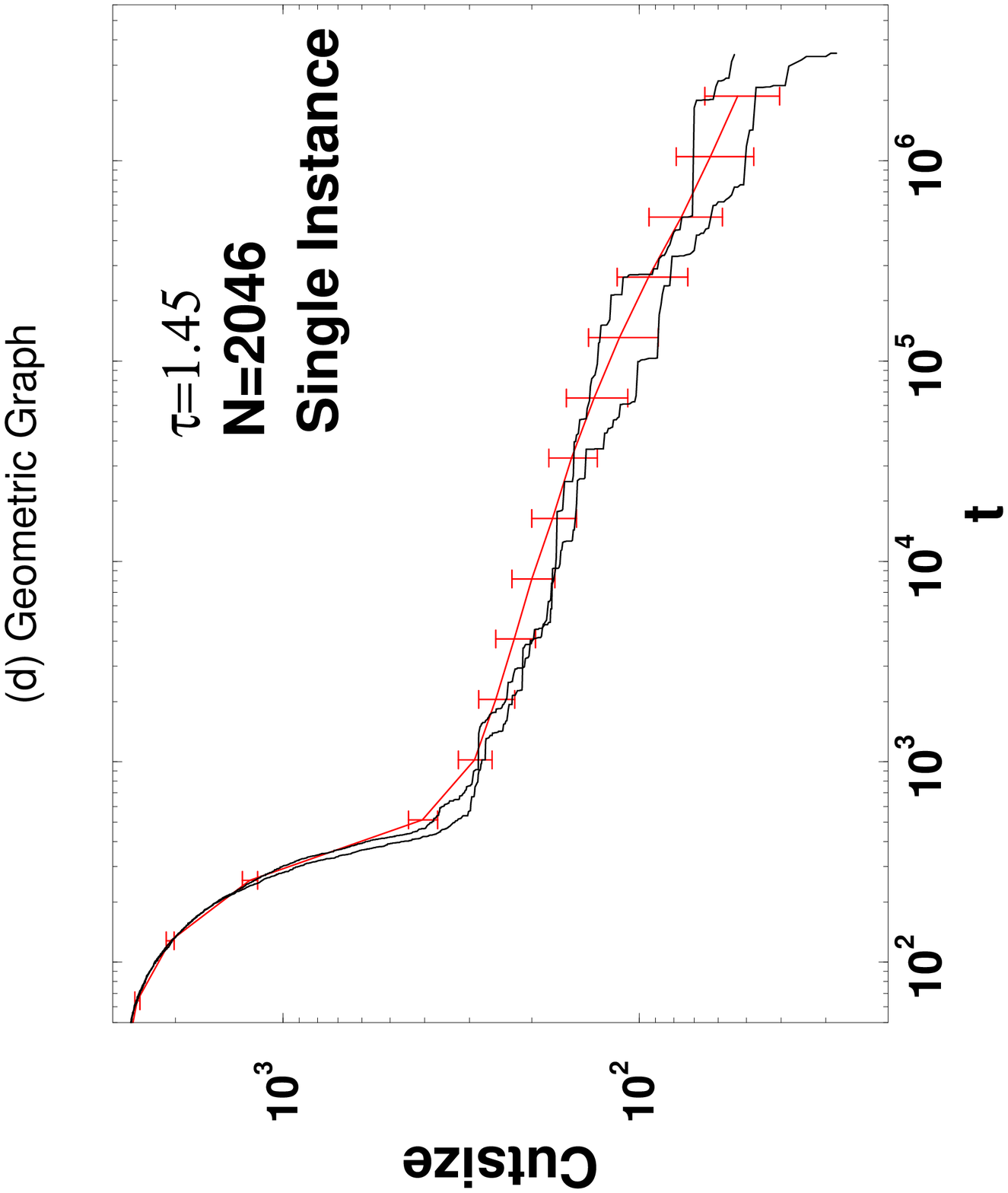}
\caption{Log-log plot of the cutsize as a function of the number of
update steps $t$ for a single (large) instance of a (a) random, (b)
trivalent, (c) ferromagnetic and (d) geometric graph.  The solid black
line represents $m(t)$ for a single run, and the solid red line with
error bars represents the average $\langle m(t)\rangle$ over all runs on
that instance.  In (a) and (b), error bars are very small, indicating
that there are only small run-to-run fluctuations about $\langle
m(t)\rangle$.  By contrast, in (c) and (d), these run-to-run
fluctuations are huge.  In fact, for the geometric graph in (d) we have
plotted two separate solid black lines representing two 
selected runs, one
poor and one good one.  This shows the large variations between
runs that lead to the large error bars.
}
\label{singleinst}
\end{figure}

As explained in Sec.~\ref{EOalgo}, the cutsize of the current
configuration will fluctuate wildly at all times during an EO run and
will not in itself converge.  Instead, we have to keep track of the {\em
best\/} configuration obtained so far during a run.  Thus, even for
times $t<t_{\rm max}$ during an EO run, we refer to the cutsize $m(t)$
of the current best configuration as the ``result'' of that run at time
$t$.  Examples of the stepwise manner in which $m(t)$ converges towards
the optimal cutsize, for a particular instance of each type of graph,
are given in Fig.~\ref{singleinst}.  Up until times $t\sim N$, $m(t)$
drops rapidly because each update rectifies poor arrangements created by
the random initial conditions.  For times $t\gg N$, it takes collective
rearrangements involving many vertices to (slowly) obtain further
improvements.

\begin{figure}[t]
\vskip 2.3truein   \includegraphics{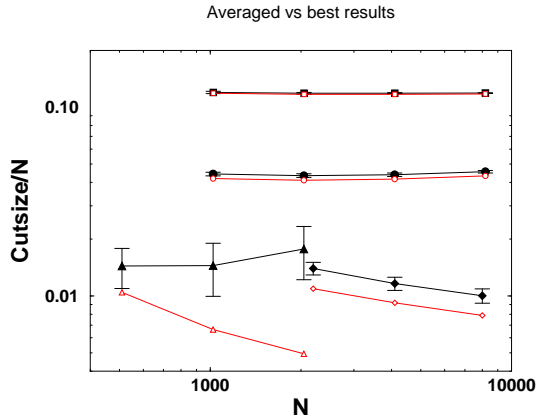}
\caption{Log-log plot of the cutsize ${\overline {\langle
m\rangle}}$, averaged over all runs and instances (filled symbol), and the best
cutsize ${\overline {m}}_{\rm best}$ from all runs on an instance, averaged
over all instances (open symbol).  This is shown for random ($\circ$),
trivalent ($\Box$),
ferromagnetic ($\diamond$), and geometric graphs ($\triangle$), as a function
of size $N$.  
The error bars refer only to run-to-run (and not instance-to-instance)
fluctuations.  For random and trivalent graphs, both average and best
cutsizes scale linearly in $N$, as expected.  For ferromagnetic and
geometric graphs, the best results are several standard deviations better 
than the average
results, with a widening gap for increasing $N$ on the geometric graphs.
The scaling of $m_{\rm best}\sim N^{1/\nu}$ gives $\nu=1.3$ for
ferromagnetic graphs, and is consistent with $\nu=2$ for geometric
graphs.}
\label{aver_best}
\end{figure}

While during each run $m(t)$ decreases discontinuously, for the random
and trivalent graphs the jumps deviate relatively little from that of
the mean behavior obtained by averaging over many runs
($\langle\ldots\rangle$).  Fluctuations, shown by the error bars in
Figs.~\ref{singleinst}a-b, thus are small and will be neglected
henceforth.  For the ferromagnetic and geometric graphs, these
fluctuations can be enormous.  In Fig.~\ref{aver_best} we compare the
average performance ${\overline {\langle m\rangle}}$ with the best
performance ${\overline {m_{\rm best}}}$ for each type of graph, at the
maximal runtime $t_{\rm max}$ and averaged over all instances at each
$N$.  The results demonstrate that for random and trivalent graphs the
average and best results are very close, whereas for ferromagnetic and
geometric graphs they are far apart (with even the scaling becoming
increasingly different for the geometric case).  Therefore, in the
following we will consider scaling of the {\em average\/} results for the first
two classes of graphs, but properties of the {\em best\/} results for the
latter two.

\subsubsection{Random and trivalent graphs}
\label{randscal}

\begin{figure}
\vskip 2.3truein   \includegraphics{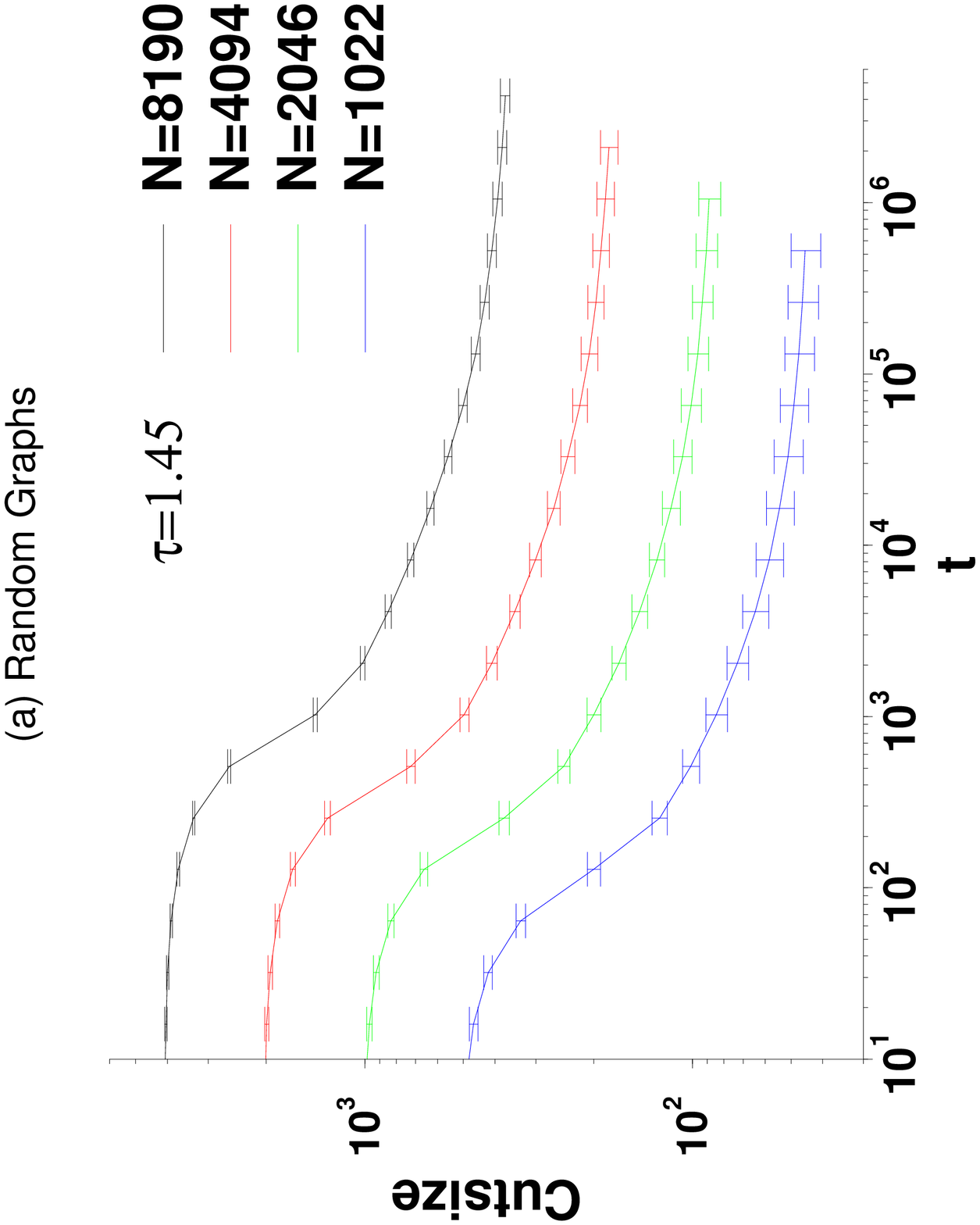}   \includegraphics{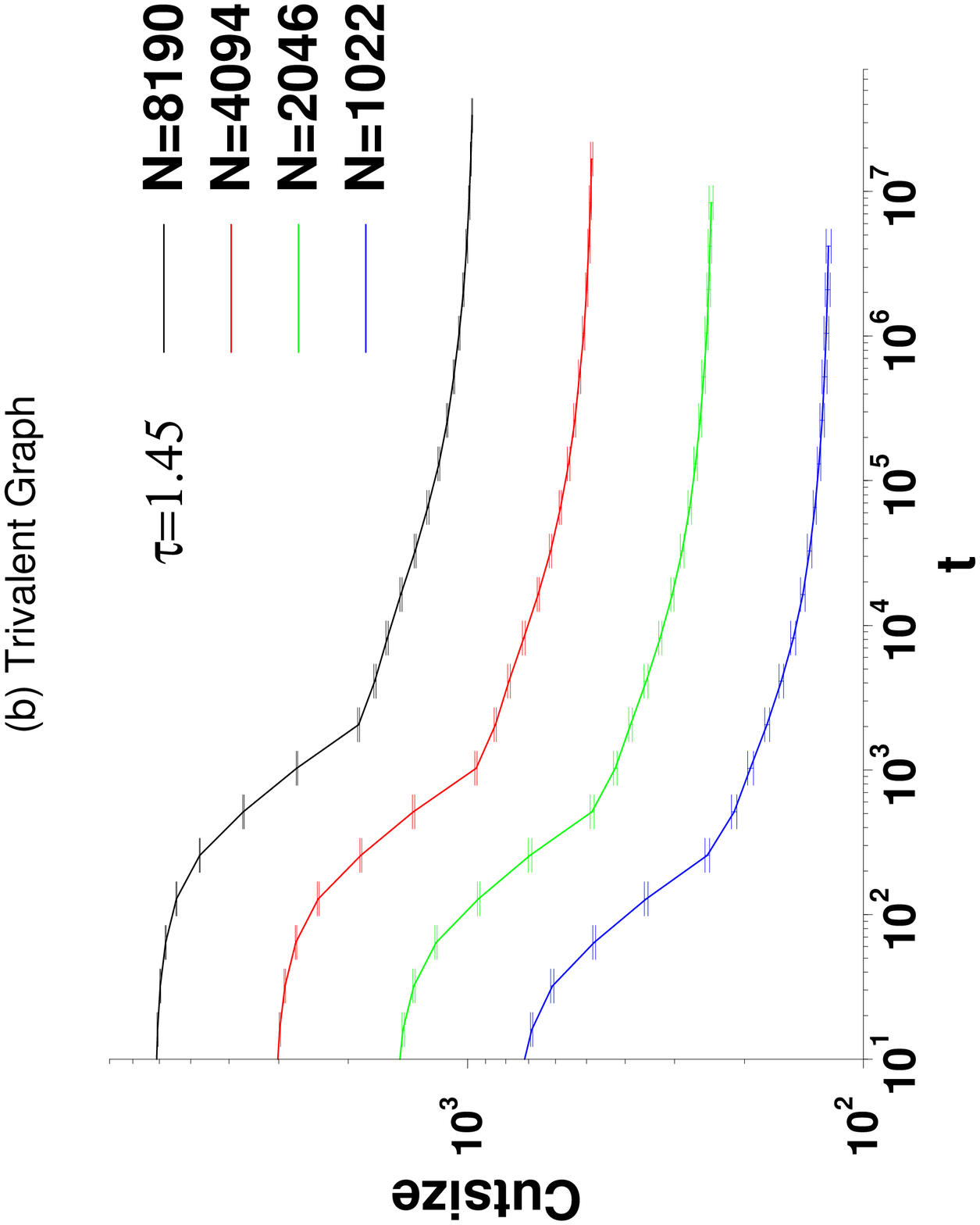}
\caption{Log-log plot of the average cutsize $m(N,t,\tau)$ as a function
of runtime $t$ at fixed $\tau=1.45$, for (a) random graphs and (b) trivalent
graphs.  The average is taken over runs as
well as over instances (32 instances for random and 8 instances for
trivalent graphs).  Error bars represent instance-to-instance
fluctuations only, and increase with $N$ much more slowly than the mean
result. In each case, $N$ increases from bottom to top.}
\label{alldata}
\end{figure}

Averaging $m(t)$ for any $t$ over all runs ($\langle\ldots\rangle$) and
over all instances (overbar), we obtain the averaged cutsize
\begin{eqnarray}
{\overline {\langle m\rangle}}=m(N,t,\tau)
\label{scalfunc}
\end{eqnarray}
as a function of size $N$, runtime $t$, and parameter $\tau$.  In
Fig.~\ref{alldata} we plot $m(N,t,\tau)$ for random and trivalent graphs
as a function of $t$, for each $N$ and at a fixed value $\tau=1.45$ that
is near the optimal value for the maximal runtimes $t_{\rm max}$ (see
Sec.~\ref{ergo}).  The error bars shown here are due only to
instance-to-instance fluctuations, and are consistent with a normal
distribution around the mean. (Note that the error bars are distorted
due to the logarithmic abscissa.)  The fact that the relative errors
decreases with $N$ demonstrate that self-averaging holds and that we
need only focus on the mean to obtain information about the typical
scaling behavior of an EO run.

\begin{figure}
\vskip 2.3truein   \includegraphics{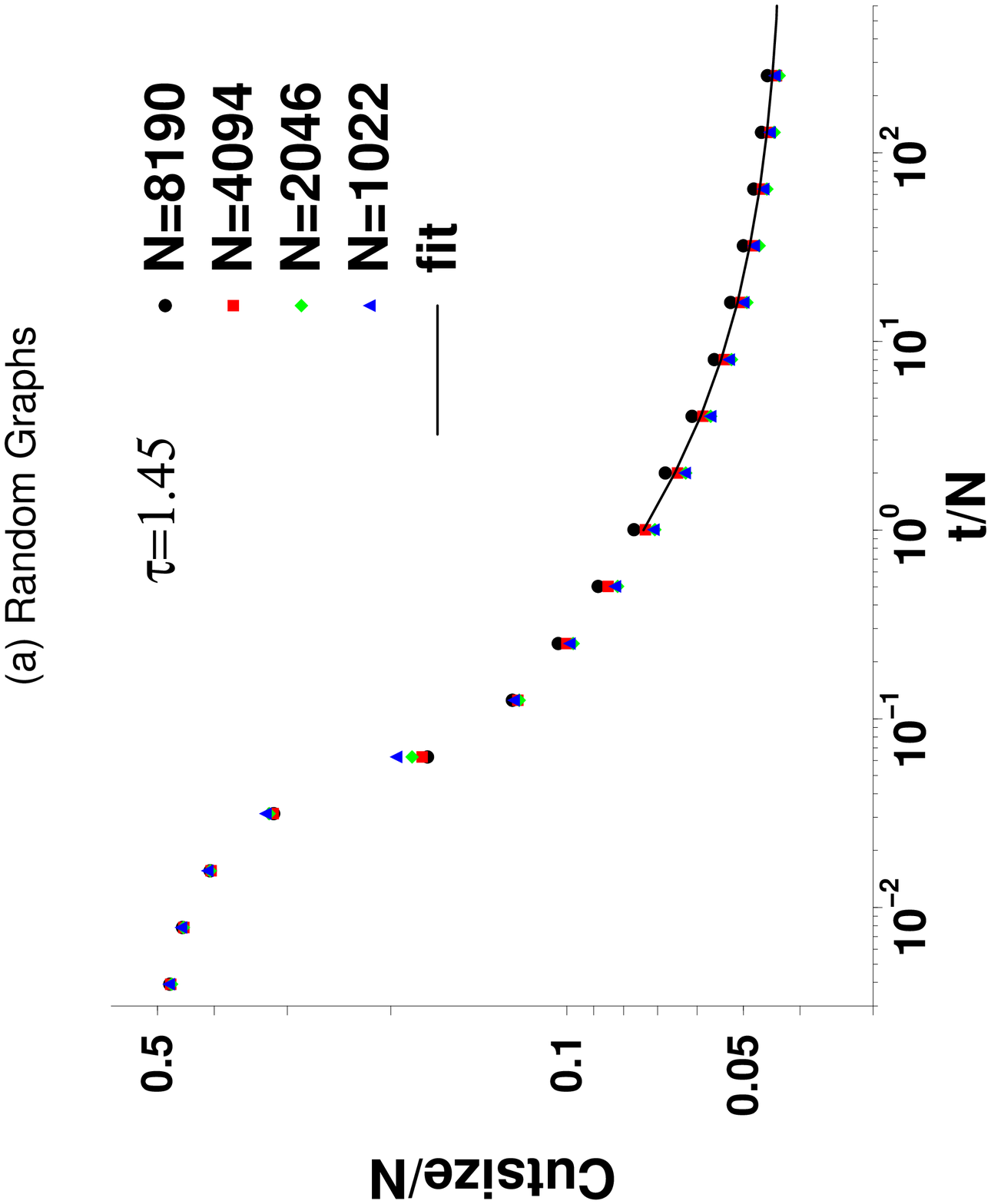}
\includegraphics{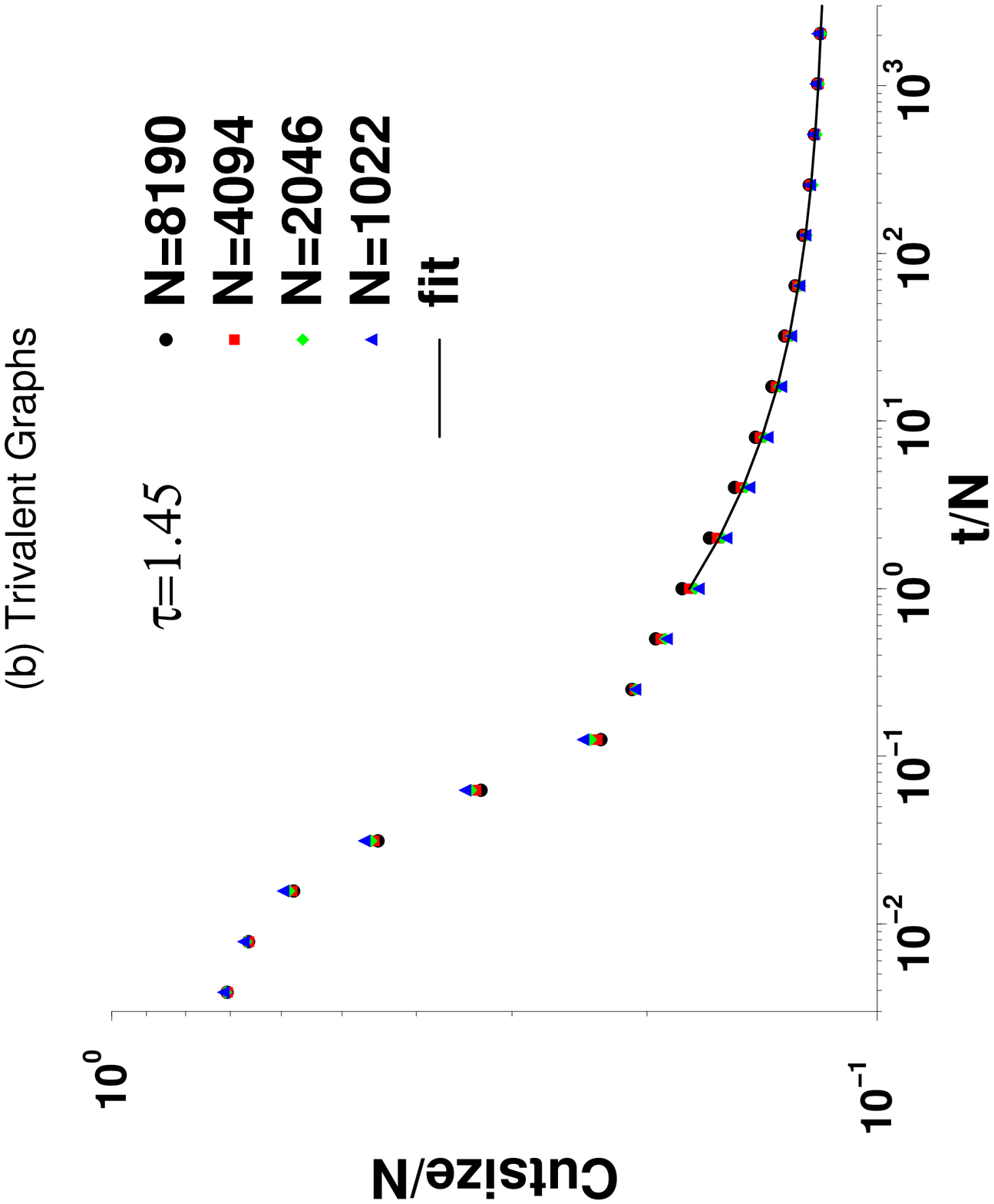}
\caption{Scaling collapse of the data from Fig.~\protect\ref{alldata}
onto a single scaling curve ${\tilde m}(t/N,\tau)=m/N$ as a function of
$t/N$ at fixed $\tau=1.45$, for (a) random graphs and (b) trivalent
graphs.  For $t>N$, data points are fitted to 
the power-law expression in
Eq.~(\protect\ref{asympteq}) with numerical values given in
Tab.~\protect\ref{fitvalues}.}
\label{collapse}
\end{figure}

We wish to study the scaling properties of the function $m$ in
Eq.~(\ref{scalfunc}).  First of all, we find that generally
\begin{eqnarray}
m(N,t,\tau)\sim N^{1/\nu} {\tilde m}(t/N,\tau)\quad (t\gg N\gg1),
\label{Nscal}
\end{eqnarray}
reflecting the fact that for EO, as well as for most other graph
partitioning heuristics that are based on local search, runtimes scale
linearly with $N$.  (This is justified by the fact that each of the $N$
variables only has 2 states to explore.  In, say, the traveling
salesperson problem, by contrast, each of $N$ cities can be reconnected
to $O(N)$ other cities, and so runtimes typically scale at least with
$N^2$~\cite{JohnsonTSP}.)  In Fig.~\ref{collapse} we plot
$m(N,t,\tau)/N$ for fixed $\tau=1.45$ as a function $t/N$. We find
indeed that the data points from Fig.~\ref{alldata} collapse onto a
single scaling curve ${\tilde m}(t/N,\tau)$, justifying the scaling
ansatz in Eq.~(\ref{Nscal}) for $\nu=1$.  The scaling collapse is a
strong indication that EO converges in $O(N)$ updates towards the optimal
result, and also that the optimal cutsize itself scales linearly in $N$.

\begin{table}[!b]%
\caption[t1]{Sequence of values of the fit of
Eq.~(\protect\ref{asympteq}) to the data in
Fig.~\protect\ref{collapse} for $t/N>1$, for each
$N$.}%
\begin{tabular}{l|c|c|c|c}
Graph Type & Size $N$ & $\langle {\tilde m}_{\rm opt}\rangle$ & $C$ &
$\gamma$ \\  \colrule   Random & 1022 & 0.0423 & 0.028 & 0.48\\ & 2046
& 0.0411 & 0.029 & 0.46\\  & 4094 & 0.0414 & 0.032 & 0.45\\  & 8190 &
0.0425 & 0.034 & 0.44\\   \colrule   Trivalent & 1022 & 0.1177 & 0.052
& 0.43\\  & 2046 & 0.1159 & 0.057 & 0.41\\  & 4094 & 0.1159 & 0.062 &
0.41\\  & 8190 & 0.1158 & 0.066 & 0.40\\ \colrule
\end{tabular}
\label{fitvalues}
\end{table}

The scaling function ${\tilde m}$ appears to converge at large times to
the average optimal result $\langle {\tilde m}_{\rm opt}\rangle=\langle
m_{\rm opt}\rangle/N$ according to a power law:
\begin{eqnarray}
{\tilde m}(t/N,\tau)\sim \langle {\tilde m}_{\rm opt}\rangle + C
\left({t\over N}\right)^{-\gamma}\quad (t\gg N\gg1).
\label{asympteq}
\end{eqnarray}
Fitting the data in Fig.~\ref{collapse} to Eq.~(\ref{asympteq}) for
$t/N>1$, we obtain for each type of graph a sequence of values for
$\langle {\tilde m}_{\rm opt}\rangle$ and $\gamma$ for increasing $N$,
given in Tab.~\ref{fitvalues}.  In both cases we find values for
$\langle {\tilde m}_{\rm opt}\rangle/N$ that are quite stable, while the
values for $\gamma$ slowly decrease with $N$.  The variation in $\gamma$
as a function of $N$ may be related to the fact that for a fixed $\tau$,
EO's performance at fixed $A=t/N$ deteriorates logarithmically with $N$,
as seen in Sec.~\ref{ergo}.  Even with this variation, however, the
values of $\gamma$ for both types of graph are remarkably similar:
$\gamma\approx 0.4$.  This implies that in general, on graphs without
geometric structure, we can halve the approximation error of EO by
increasing runtime by a factor of 5--6.  This power-law convergence of
EO is a marked improvement over the mere logarithmic convergence
conjectured for simulated annealing~\cite{Science} on NP-hard
problems~\cite{Grest}.

Finally, the asymptotic value of $\langle {\tilde m}_{\rm opt}\rangle$
obtained for the trivalent graphs can be compared with previous
simulations~\cite{Banavar,EOperc}.  There, the ``energy'' ${\cal
E}=-1+4\langle {\tilde m}_{\rm opt}\rangle/3$ was calculated using the
best results obtained for a set of trivalent graphs.
Ref.~\cite{Banavar} using simulated annealing obtained ${\cal
E}=-0.840$, while in a previous study with EO~\cite{EOperc} we obtained
${\cal E}=-0.844(1)$. Our current, somewhat more careful extrapolation
yields $\langle {\tilde m}_{\rm opt}\rangle= 0.1158N$ or ${\cal
E}=-0.845(1)$.  Even though this extrapolation is based on the average
data rather than only the best of all runs, the very low fluctuations
between runs (see Figs.~\ref{singleinst}b and \ref{aver_best}) indicate
that the result for ${\cal E}$ would not change significantly.  Thus,
the replica symmetric solution proposed in Refs.~\cite{M+P,W+S} for this
version of the GBP, which gives a value of ${\cal E}=-2\times
0.7378/\sqrt{3}=0.852$, seems to be excluded.

\subsubsection{Ferromagnetic and geometric graphs} 
\label{geomscal}
Unlike on the preceding graphs, EO gives significantly different results
for an average run and for the best run on geometrically structured
graphs (see Figs.~\ref{singleinst} and \ref{aver_best}). In
Fig.~\ref{aver_best}, at least the results from the best run (averaged
over all instances) comes close to
the scaling behavior expected from the considerations in
Sec.~\ref{scaling}: a fit gives $\nu\approx1.3$ for ferromagnetic, and
$\nu\approx2$ for geometric graphs, while the theory predicts $\nu=3/2$ and
$\nu=2$ respectively. But even these best cutsizes themselves vary
significantly from one
instance to another.  Thus, it is certainly not useful to study the
``average'' run.
Instead, we will consider the result of each run at the maximal runtime
$t_{\rm max}$, extract the best out of $k$ runs, and study these results
as a function of increasing $k$.

\begin{figure}
\vskip 2.3truein \includegraphics{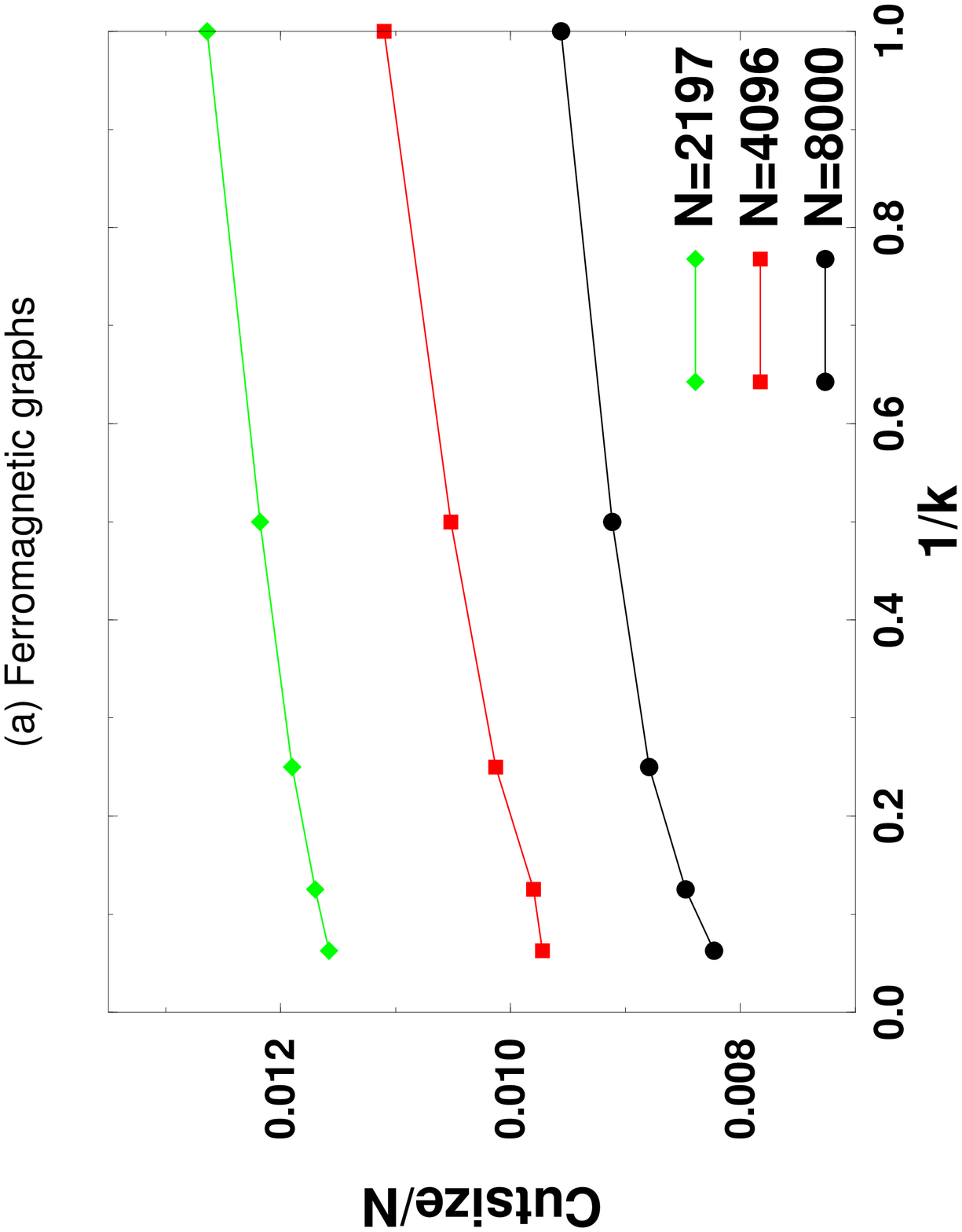} \includegraphics{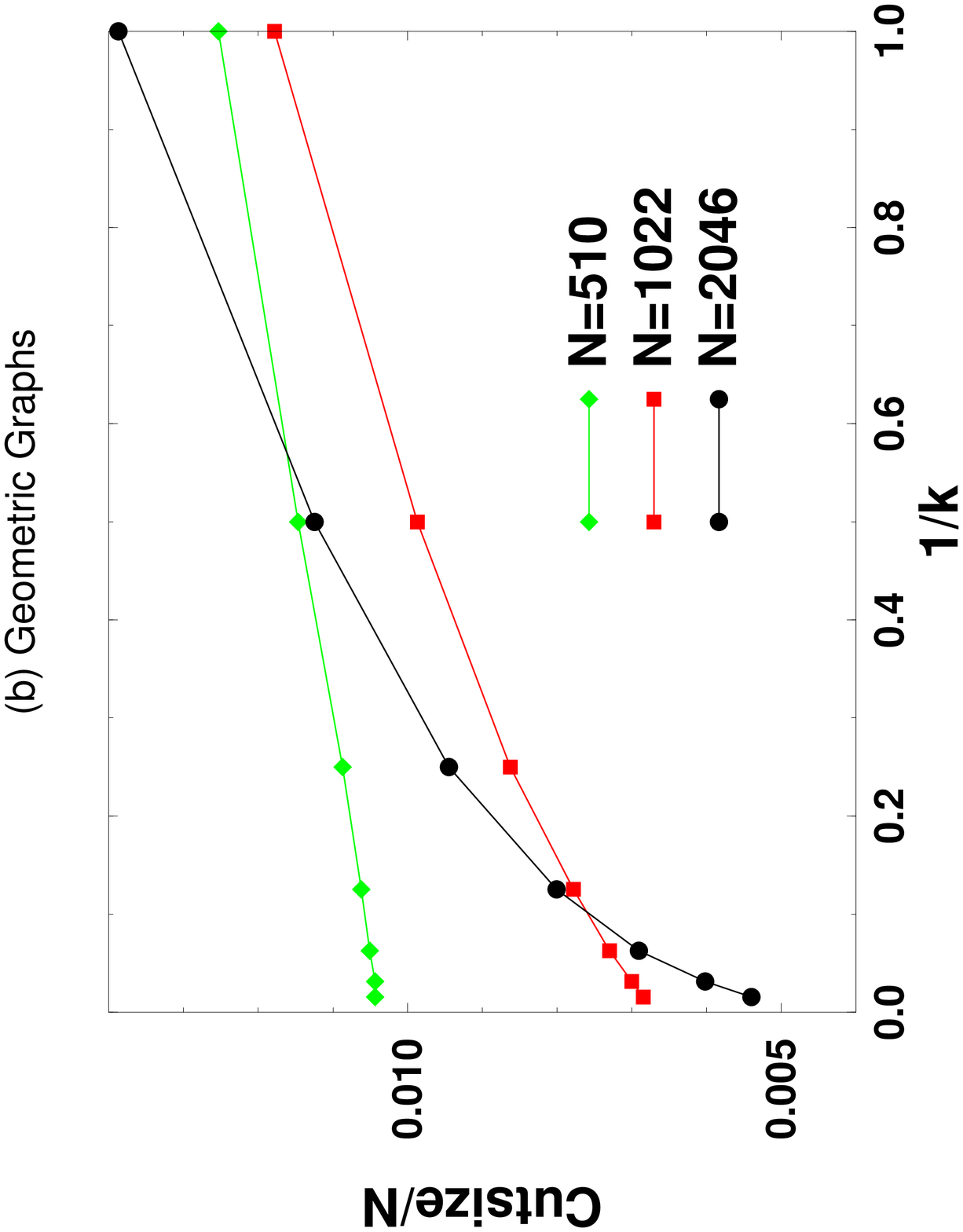}
\caption{Extrapolation plot for the best-of-$k$ trials for (a)
ferromagnetic graphs at $\tau=1.4$ and (b) for geometric graphs at
$\tau=1.3$.  The data for this plot are extracted from the results at
$t_{\rm max}$, averaging the best-of-$k$ results over 16 instances for
ferromagnetic, and 64 instances for geometric graphs.  For comparison,
the leftmost data point for each $N$ at $k=16$ for ferromagnetic graphs
and at $k=64$ for geometric graphs corresponds to the ``best'' results
plotted in Fig.~\protect\ref{aver_best} for those graphs.}
\label{bestofk}
\end{figure}

Fig.~\ref{bestofk} shows the difficulty of finding good runs with
near-optimal results for increasing size $N$.  While for ferromagnetic
graphs it is possible that acceptable results may be obtained without
increasing $k$ dramatically, for geometric graphs an ever larger number
of runs seems to be needed at large $N$.  It does not appear possible to
obtain consistent good near-optimal results at a fixed $k$ for
increasing $N$.

\begin{figure}
\vskip 2.3truein
\includegraphics{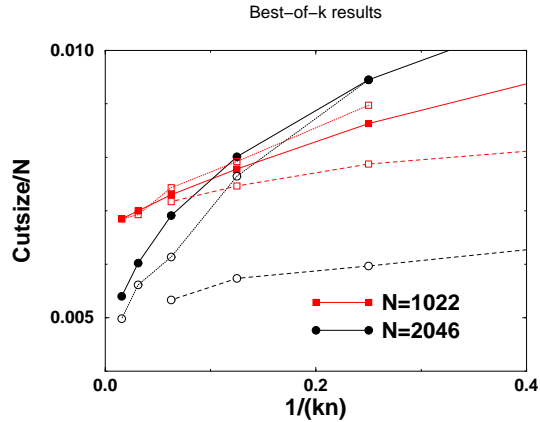}
\caption{Equivalent-runtime comparison between different strategies to
improve EO results on geometric graphs for $N=1022$ and 2046 at
$\tau=1.3$. The horizontal axis is proportional to the inverse of the
number of updates used, $t=k\times n\times t_{\rm max}$.  Filled symbols
refer to the $n=1$ results for geometric graphs already plotted in
Fig.~\protect\ref{bestofk}b, where $k$ is varied.  Open symbols on the
dotted lines refer to the $k=4$ (best-of-4) results, where $n$ is
varied.  Opaque symbols on the dashed line refer to $k=4$ results as
well, but using initial conditions generated by a clustering algorithm (see
Sec.~\protect\ref{startup}).
At sufficiently long runtime all strategies are about equal, though with
fewer but longer runs having a slight edge over more and shorter runs
for large $N$.  Even the advantages of a non-random initial configuration
become less significant at longer runtimes.}
\label{bestof_compare}
\end{figure}

We saw in Sec.~\ref{randscal} that for random graphs, computational time
is well spent on a few, long EO runs per instance.  We can not address
in detail the question of whether, for geometrically defined graphs,
computational time is better spent on $k$ independent EO-runs with
$t_{\rm max}$ update steps or, say, on a single EO-run with $k\times
t_{\rm max}$ update steps.  While experience with our data would
indicate the former to be favorable, an answer to this question depends
significantly on $N$ and, of course, on the choice of $\tau$ (see
Sec.~\ref{ergo}).  Here we consider this question merely for a single
value, $\tau=1.3$, for which we have run EO on the same 64 geometric
graphs up to 16 times longer than $t_{\rm max}$, but using only $k=4$
restarts.  In each of these 4 runs on an instance we recorded the best
result seen at multiples $n\times t_{\rm max}$ with $n=1,2,4,8$, and 16.
For example, the best-of-4 runs at $n=1$ of this runtime corresponds to
the best-of-$k$ results in Fig.~\ref{bestofk} for $k=4$, while $n=16$
would correspond to the same amount of running time as $k=64$.
Fig.~\ref{bestof_compare} shows that fewer but longer runs are slightly
more successful for larger $N$. 

Finally, we have also used the clustering algorithm described in
Sec.~\ref{startup} and Ref.~\cite{BoPe1} on this set of 64 graphs with
$\tau=1.3$.  For comparison, we again use the best-of-4 runs with
averages taken at times $n\times t_{\rm max}$, $n=1,2,$ and 4.  Results
at short runtimes improve by a huge amount with such a procedure, but
its advantage is eventually lost at longer runtimes.  While this
procedure is cheap, easy, and apparently always successful for geometric
graphs, our experiments indicate that its effect may be less significant
for random graphs and may actually result in {\em diminished\/}
performance when used for trivalent graphs in place of random initial
conditions.  Clearly, clustering is tailored more toward geometric
graphs satisfying a triangular inequality.

\section{Conclusions}
\label{Conclusion}
Using the classic combinatorial optimization problem of bipartitioning
graphs as an application, we have demonstrated a variety of properties
of the extremal optimization algorithm.  We have shown that for random
graphs, EO efficiently approximates the optimal solution even at large
size $N$, with an average approximation error decreasing over runtime
$t$ as $t^{-0.4}$.  For sparse, geometrically defined graphs, finding
the ideal (sub-dimensional) interface partitioning the graph becomes
ever more difficult as size $N$ increases.  EO, like other local search
methods~\cite{JohnsonGBP}, gets stuck ever more often in poor local
minima.  However, when we consider the best out of multiple runs with
the EO algorithm, we recover results close to those predicted by
theoretical arguments.

We believe that many of our findings here, notably with regard to EO's
fitness definition and the update procedure using the parameter $\tau$,
are generic for the algorithm.  Our results for optimizing 3-coloring
and spin glasses appear to bear out such generalizations~\cite{EO_PRL}.
In view of this observation, a firmer theoretical understanding of our
numerical findings would be greatly desirable.  The nature of EO's
performance derives from its large fluctuations; the price we pay for
this is the loss of detailed balance, which is the theoretical basis for
other physically-inspired heuristics, such as simulated
annealing~\cite{Science}.  On the other hand, unlike in simulated
annealing, we have the advantage of dealing with a Markov chain that is
{\em independent\/} of time~\cite{Aarts}.  This suggests that our method
may indeed be amenable to theoretical analysis.  We believe that the
results EO delivers, as a simple and novel approach to optimization,
justify the need for further analysis.

\section{Acknowledgements}
We would like to thank the participants of the Telluride Summer workshop
on complex landscapes, in particular Paolo Sibani, for fruitful
discussions, and Jesper Dall for confirming many of our results in his
master thesis at Odense University.  This work was supported by the
University Research Committee at Emory University, and by an LDRD grant
from Los Alamos National Laboratory.

\end{document}